\documentclass[11pt,preprint]{aastex}

\usepackage[dvips]{color}
\usepackage{amsmath}
\usepackage{morefloats}

\shorttitle{Prominence Support by Magnetic Dips}
\shortauthors{A. Hillier \& A. van Ballegooijen}

\begin{document}
\title{On the Support of Solar Prominence Material by the Dips of a Coronal Flux Tube} 

\author{Andrew Hillier}
\affil{Kwasan and Hida Observatories, Kyoto University}

\and

\author{Adriaan van Ballegooijen} 
\affil{Harvard-Smithsonian Center for Astrophysics, Cambridge, MA 02138, USA}

\begin{abstract}

The dense prominence material is believed to be supported against gravity through the magnetic tension of dipped coronal magnetic field.
For quiescent prominences, which exhibit many gravity-driven flows, hydrodynamic forces are likely to play an important role in the determination of both the large and small scale magnetic field distributions.
In this study, we present the first steps toward creating three-dimensional magneto-hydrostatic prominence model where the prominence is formed in the dips of a coronal flux tube.
Here 2.5D equilibria are created by adding mass to an initially force-free magnetic field, then performing a secondary magnetohydrodynamic relaxation.
Two inverse polarity magnetic field configurations are studied in detail, a simple o-point configuration with a ratio of the horizontal field ($B_x$) to the axial field ($B_y$) of $1$:$2$ and a more complex model that also has an x-point with a ratio of $1$:$11$.
The models show that support against gravity is either by total pressure or tension, with only tension support resembling observed quiescent prominences.
The o-point of the coronal flux tube was pulled down by the prominence material, leading to compression of the magnetic field at the base of the prominence.
Therefore tension support comes from the small curvature of the compressed magnetic field at the bottom and the larger curvature of the stretched magnetic field at the top of the prominence.
It was found that this method does not guarantee convergence to a prominence-like equilibrium in the case where an x-point exists below the prominence flux tube.
The results imply that a plasma $\beta$ of $\sim 0.1$ is necessary to support prominences through magnetic tension.

\end{abstract}

\section{Introduction}

Quiescent prominences/filaments are large structures made of relatively cool plasma, which exist in quiet-sun regions of the corona. 
Prominences, observed in chromospheric lines, have a temperature of approximately $10^4$\,K \citep{TH1995} and number density $3$\,--\,$6 \times 10^{11}$\,cm$^{-3}$\citep{HIR1986}, which gives a density of $\sim 10^{-13}$ g cm$^{-3}$ giving a decrease and increase of approximately two orders of magnitude respectively from the surrounding corona. 
The pressure scale height of a prominence ($\Lambda $) can be calculated to be $\Lambda \sim 300$\,km, which is about two orders of magnitude less than the characteristic height of observed quiescent prominences ($\sim 25$\,--\,$50$\,Mm).
Using a characteristic gas pressure of $0.6$\,dyn\,cm$^{-2}$ \citep{HIR1986} and magnetic field of $3$\,--\,$30$ G \citep{LER1989} the plasma $\beta$ (ratio of gas pressure to magnetic pressure) of a quiescent prominence can be estimated as $\beta \sim 0.01$\,--\,$1$. 
For reviews on the current understanding of solar prominences see, for example, \citet{LAB2010, MAC2010}.

Globally, quiescent prominences are incredibly stable and often exist in the corona for weeks.
On smaller scales, however, the  prominence/filament system displays many localised instabilities and flows.
Observations of quiescent prominences have shown downflows \citep{ENG1981}, plumes \citep{STWI1973}, vortices of approximately $10^5$ km $\times 10^5$\,km in size \citep{LZ1984} and rising bubbles \citep{DT2008} with velocities of 10-30\,km s$^{-1}$.
The launch of the \textit{Hinode} satellite revealed lots of flows orientated along the direction of gravity as well as many vertical prominence threads \citep{BERG2008,BERG2010,Chae2010,VB2010,HILL2011a}, which implies that gravity must be an important force in the quiescent prominence system.

It has long been hypothesised that the support of prominence material against gravity is by the magnetic tension of a curved magnetic field \citep{KS1957,KR1974}.
In this way, it is possible to maintain a prominence that is significantly taller than the pressure scale height.
The Kippenhahn-Schl\"{u}ter model is one such prominence model, where Lorentz force, which results from the curvature of the magnetic field, gives sufficient magnetic tension to balance the gravitational force of the dense plasma. 
This model describes the local structure of the prominence and, as a result, has no transition to a hot corona and is infinite in extent in the vertical direction.
This model has been shown to be both linearly and nonlinearly stable to ideal Lagrangian magnetohyrodynamic (MHD) perturbations \citep{KS1957,AN1969,ZWEIB1982,GAL1984,ALY2011} .

There have been a number of recent developments in terms of the development of prominence models on larger scales.
Complex 2.5D equilibria that include a large-scale magnetic field and corona have been studied by \citet{HOOD1990}, \citet{PET2007} and \citet{BLOK2011}.
These studies have managed to reproduce many observed prominence features, including the formation of double layered prominences.
Recent studies on coronal condensation have shown that this is a very promising mechanism to form prominences self-consistently in a magnetised corona \citep{LUNA2012, XIA2012}.

The most common approach for modelling observed prominences/filaments is that of directly modelling the magnetic field of prominence/filament systems using observations of the photospheric magnetic field \citep[e.g.][]{AUL2003, DUD2008, SVB2012}.
However, straight extrapolations of the photospheric magnetic field using either potential or nonlinear force-free field (NLFFF) extrapolations often fail to create the dips in the magnetic field that are necessary to support plasma.
To circumnavigate this issue, the most common method is to artificially insert a flux rope into the coronal magnetic field and calculate a new equilibrium.

\citet{AUL2003} presented models using photospheric magnetic field measurements to calculate the coronal magnetic field around observed filaments.
To these magnetic field extrapolations, flux tubes were added, and the positions of the dips in the flux tubes were compared with the filament/prominence structure.
The results showed that the average field strength in the quiescent prominence modelled was approximately $3$\,G.
It was also found that the positions of the dips reproduced the global structure of the prominence/filament reasonably well. 

In an attempt to match the observations of the vertical thread structure of quiescent prominences with theoretical expectations, \citet{VB2010} proposed a model in which the threads are supported by a tangled magnetic field. 
As in classic prominence models, the Lorentz force from dips in the magnetic field supports the plasma against gravity. 
The vertical threads are hypothesized to form as a result of magnetic Rayleigh--Taylor instability in the tangled field. 
However, it is not clear whether such tangled fields indeed exist in prominences, and if so, how the tangled field is formed. 
Therefore, in the present paper we return to a more standard scenario in which the prominence plasma is located at the dips of a large-scale magnetic flux rope  \citep{KR1974, PN1983, PR1989, RK1994, LH1995, AUL1998, GIB2006, DUD2008, SVB2012}.

In this paper we perform 2.5D numerical simulations to investigate the prominence structure obtained by inserting mass into a flux tube in the solar corona.
The aim of this work is understand how the addition of prominence material alters the structure of the coronal magnetic field and to investigate the equilibrium formed.

\section{Numerical Method and Initial Setting}\label{SETTING}

In this section the basic assumptions of the prominence modelling are described. 
A quiescent prominence and its local environment are considered.  
For simplicity we use a Cartesian reference frame $(x,y,z)$, where $x$ is the horizontal coordinate perpendicular to the prominence axis, $y$ is the coordinate along the long axis of the prominence, and $z$ is the height above the photosphere. 
The coordinates are expressed in units of the pressure scale height $\Lambda$ of the coronal plasma ($\Lambda \approx 55$ Mm for a 1 MK corona). 
The $x$ coordinate is in the range $-L \le x \le L$, where $L$ is the half-width of the computational domain ($L = 2.5\Lambda$). 
The magnetic field and plasma parameters are assumed to be independent of the $y$ coordinate, so the models are 2.5D.

The construction of the models proceeds in two steps. 
First, an appropriate magnetic field able to support the prominence plasma is constructed. 
These initial fields are assumed to be NLFFFs containing dips in the magnetic field lines. 
Then the prominence plasma is inserted into the models at the dips, and the system is evolved using a time-dependent MHD code until an equilibrium is reached.  
The typical height of a quiescent prominence ($50$\,Mm) is comparable to the pressure scale height $\Lambda$ of the corona plasma, so the modelling must take into account the effects of gravity, not only in the prominence but also in the surrounding corona.  
In the following we first describe the construction of the NLFFF models, and then discuss the numerical methods used in the MHD simulations.

Solar prominences can be classified as "normal" or "inverse" polarity
depending on the direction of magnetic field at the prominence when compared to the potential field. A normal polarity prominence model was recently developed by \citet{XIA2012}. In this model the prominence is formed by evaporation of plasma from the chromosphere and condensation at the top of a sheared arcade. The configuration is assumed to be symmetrical with respect to the mid-plane of the arcade. We suggest that the stability of the prominence may be strongly affected by the symmetry
of the configuration. If the magnetic field and/or plasma flows were asymmetric, the prominence would be pushed to one side of the arcade, as has been found in one-dimensional asymmetric loop models \citep[e.g., see Figure 11 in ][]{XIA2011}. Also, the prominence may be subject to a gravitational instability that distorts the magnetic field lines, causing the prominence to fall down along one side. It has not yet been demonstrated that normal polarity prominences can be stable to such perturbations. Therefore, in the present paper we focus on inverse polarity prominences containing flux ropes, which are believed to be more stable to such gravity-driven instabilities.

\subsection{Construction of Nonlinear Force-free Field}

The initial configuration is assumed to be a NLFFF containing a twisted magnetic flux rope. 
The axis of the rope lies horizontally above the polarity inversion line in the photosphere (i.e, the line $x= z = 0$). 
The flux rope is held down by an overlying coronal arcade that is anchored in the photosphere on the side of the prominence. 
The invariance of the magnetic field ${\bf B}$ with respect to $y$ implies
\begin{equation}
B_x = - \frac{\partial A}{\partial z} , ~~~~~ 
B_z = \frac{\partial A}{\partial x} ,
\label{BxBz}
\end{equation}
where $A(x,z)$ is the magnetic flux function. 
Note that the contours of $A(x,z)$ are the projections of field lines onto the $x$--$z$ plane. 
Inserting the above expression into the force-free condition, $( \nabla \times {\bf B}) \times {\bf B} = 0$, we find that $B_y$ is a function of $A$, which we write as $B_y (x,z) \equiv \tilde{B} [A(x,z)]$. 
Also, $A(x,z)$ must satisfy the following partial differential equation:
\begin{equation}
\nabla_\perp^2 A + \tilde{B} \frac{d \tilde{B}} {dA} = 0 ,
\label{eq:D2A}
\end{equation}
where $\nabla_\perp$ is the derivative in the $x$--$z$ plane. 
Using Ampere's law, we see that the first term is closely related to the $y$-component of the electric current density, $j_y$:
\begin{equation}
- \nabla_\perp^2 A = ( \nabla \times {\bf B} )_y = \frac{4 \pi}{c}
j_y (x,z) ,
\end{equation}
where $c$ is the speed of light. 
At the top and side boundaries of the computational domain we use $A = 0$.

Two different NLFFF models are considered. 
The flux distributions on the photosphere are different for the two models:
\begin{eqnarray}
A(x,0) & = & A_0 \cos ( \onehalf \pi x / L )  ~~~
\mbox{for model 1} , \\ 
A(x,0) & = & A_0 / \left[ 1 + (x/x_0)^4 \right] ~~~
\mbox{for model 2} ,
\end{eqnarray}
where $A_0$ is a constant (we use $A_0 = 100 \Lambda$), and $x_0$ is the half-width of the flux distribution for case 2 ($x_0 = 0.5 \Lambda$). 
The flux distribution $B_z (x,0)$ can be obtained by taking the derivative of $A(x,0)$ with respect to $x$. 
The axial magnetic fields are also different for the two models:
\begin{eqnarray}
\tilde{B} (A) & = &  \frac{C_0 A_0} {\Lambda} \left[ 1 - \exp \left(
\frac{A} {2 A_0} \right) \right]^2 ~~~~ \mbox{for model 1} , 
\label{eq:B1} \\
\tilde{B} (A) & = &  \frac{C_0 A_0} {\Lambda} \left\{ 1 - \exp \left[
- \left( \frac{A} {0.6 A_0} \right)^3 \right] \right\} ~~~~
\mbox{for model 2} \label{eq:B2} ,
\end{eqnarray}
where $C_0$ is a measure of the deviation from the potential field.
If $C_0$ is sufficiently large, the solution $A(x,z)$ of Equation (\ref{eq:D2A}) will contain a local maximum, which corresponds to the axis of a twisted flux rope. Expression (\ref{eq:B1}) increases monotonically with $A$, while Equation (\ref{eq:B2}) saturates for large $A$ values. 
This leads to significant differences in the distribution of electric currents in and around the flux rope. 
Model 1 was designed such that the flux rope rests on the photosphere, while in model 2 there is an X-line between the photosphere and the flux rope.

Together with the boundary conditions, Equation (\ref{eq:D2A}) represents a nonlinear boundary-value problem that must be solved by iteration. 
Let $A_k (x,z)$ be the flux function for iteration $k$, then we can compute $B_k = \tilde{B} (A_k)$ and $\nabla_\perp^2 A_{k+1} \approx -B_k dB_k / dA_k \equiv - C_k(x,z)$. 
The field $A_{k+1} (x,z)$ is written as a sum of potential and non-potential components, $A_{k+1} = A_{pot} + \delta A_{k+1}$. 
The potential field satisfies $\nabla_\perp^2 A_{pot} = 0$ and is uniquely determined from the boundary conditions at the photosphere (see Equations (4) and (5)). 
The non-potential field satisfies $\nabla_\perp^2 ( \delta A_{k+1} ) = - C_k (x,z)$ with boundary conditions $\delta A_{k+1} = 0$ at all four boundaries of the computational domain. This equation can be solved using a Fourier method. 

\begin{figure*}[ht]
  \begin{center}
\includegraphics[width=7.5cm]{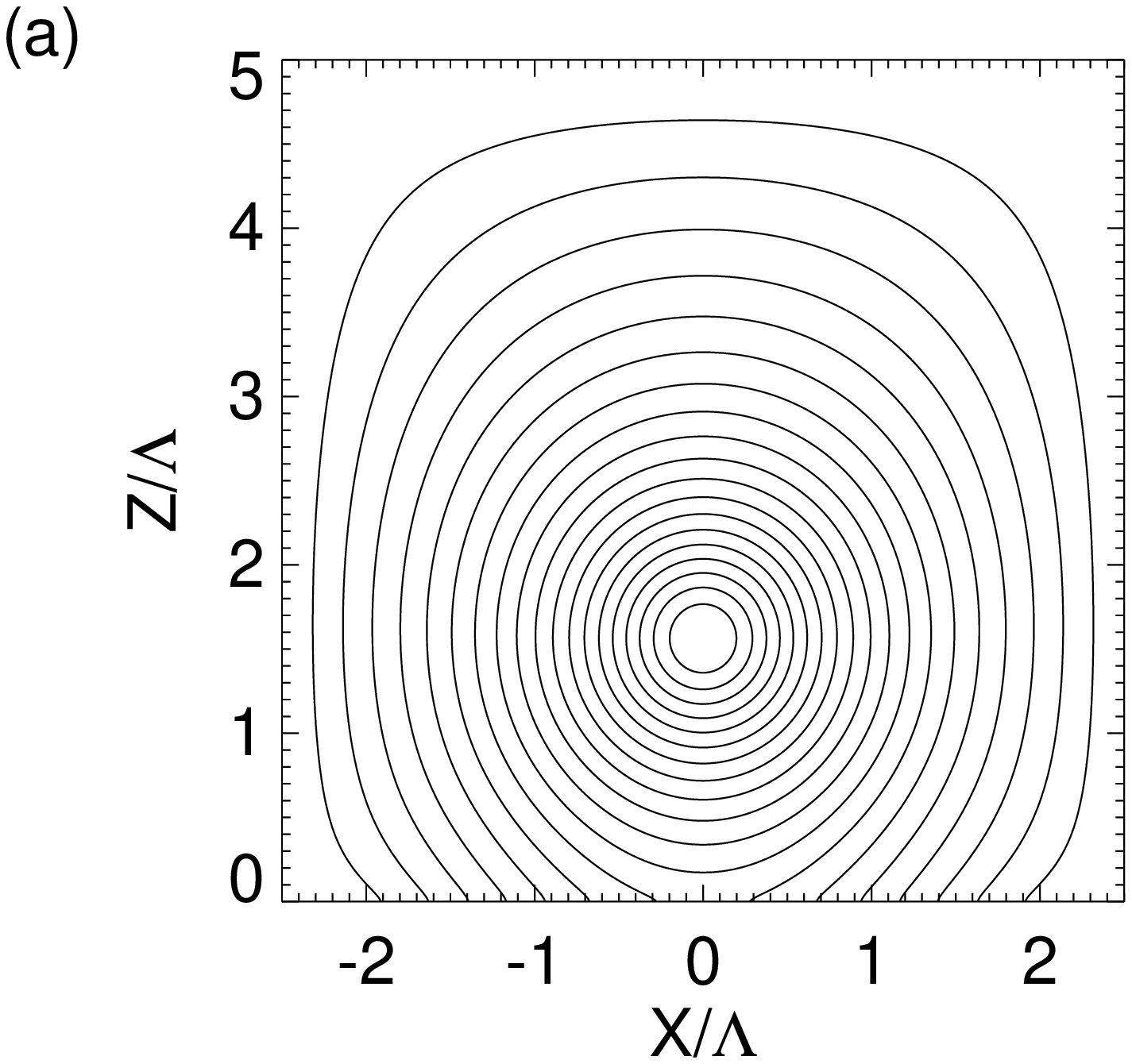}
\includegraphics[width=7.5cm]{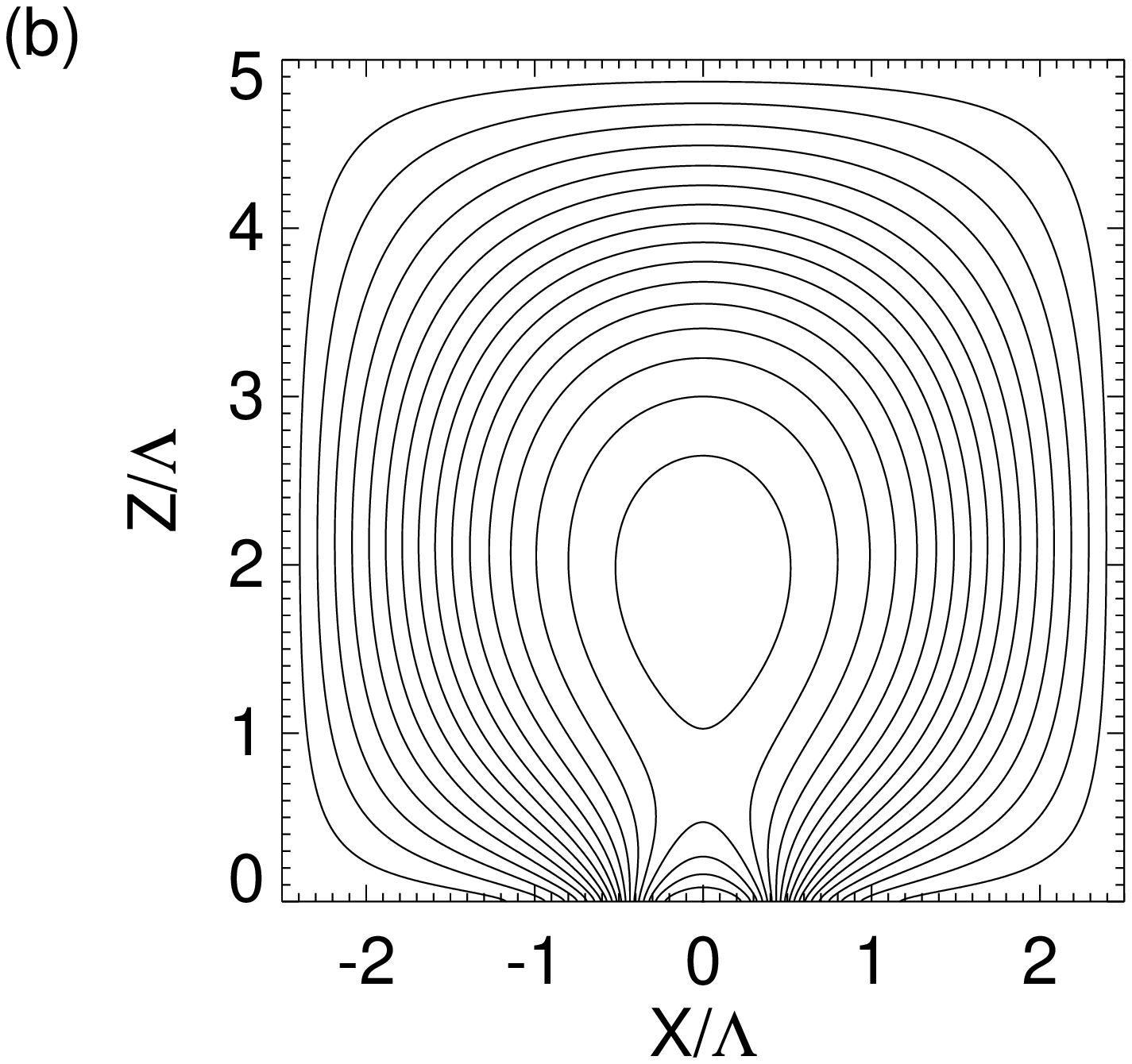}
  \end{center}
  \caption{Initial distribution of the magnetic field for (a) model 1 and (b) model 2. Contours show the magnetic vector potential.}
\label{model}
\end{figure*}

First, the domain size is quadrupled by mirroring $C_k (x,z)$ with respect to both the upper and right boundaries, and reversing the sign of $C_k$. 
The functions $C_k (x,z)$ and $\delta A_{k+1} (x,z)$ are written as Fourier series on this enlarged domain. 
Then the mode amplitudes of $\delta A_{k+1}$ are given by $\tilde{A}(k_x,k_z) = \tilde{C}(k_x,k_z) / (k_x^2+k_z^2)$, where $k_x$ and $k_z$ are wavenumbers and $\tilde{C}(k_x,k_z)$ are mode amplitudes of $C_k(x,z)$. 
This yields the next approximation of the flux function, $A_{k+1} (x,z)$. 
The parameter $C_0$ is updated in each iteration such that the system has a prescribed total electric current, $J_y \equiv \int \int j_y (x,z) dx dz$, integrated over the whole domain. 
We note that the topology of the magnetic field is not conserved, therefore, X-points may appear and disappear during this iteration process. 
Since the total current $J_y$ is fixed, the iteration always converges to a solution (generally in a few hundred iterations or less). 
The final NLFFF solution depends on the assumed value of the total current. 
Expressed in units of $A_0 c / (4 \pi)$, $J_y$ equals 837.8 and 544.5 for cases 1 and 2, respectively. 
The two models are shown in Figure \ref{model}.
It should be noted that the 2.5D flux ropes considered in this work inherently do not consider the anchoring of the flux tube ends in the photosphere.

\citet{BLOK2011} obtained prominence models by solving the magnetostatic problem for cylindrical flux ropes under the assumption that either the temperature, density or entropy is a flux function. 
A similar approach could have been used here. However, we find that when the weight of the prominence plasma is large compared to the field strength an equilibrium solution may not exist. 
Our chosen approach of inserting mass into a 2.5D model of the prominence is more flexible in dealing with such non-equilibrium cases. 
Also, in future we intend to apply our methods to full three-dimensional prominence simulations, in which case the iterative method for finding magnetostatic equilibria cannot be used. 

\subsection{Numerical Method for Simulations}

In this study, we use the ideal MHD equations. 
Constant gravitational acceleration is assumed, but viscosity, heat conduction and radiative cooling terms are neglected. 
The equations are expressed as follows:

\begin{eqnarray}\label{cont}
\frac{\partial \rho}{\partial t} + \nabla \cdot (\rho \mathbf v)=S_{\rho} \\
\frac{\partial \rho \mathbf v}{\partial t} +\nabla \cdot \left( \rho \mathbf v \mathbf v +p \mathbf I - \frac{\mathbf B \mathbf B}{4 \pi}+\frac{\mathbf B ^2}{8 \pi} \mathbf{I} \right) =\rho \mathbf g+S_{Mom}  \\
\frac{\partial \mathbf B }{\partial t}=\nabla \times (\mathbf v \times \mathbf B)\\
\frac{\partial}{\partial t} \left(\epsilon + \frac{\mathbf B ^2}{8 \pi}  \right) + \nabla \cdot \left[ (\epsilon + p) \mathbf v +\frac{c}{4\pi} \mathbf E \times \mathbf B \right ]=\rho \mathbf g \cdot \mathbf v+S_{En}\\
\mathbf E =-\frac{1}{c} \mathbf v \times \mathbf B\\
\epsilon = \frac{1}{2}\rho v^2 + \frac{p}{\gamma -1}
\end{eqnarray}
where $U$ is the internal energy per unit mass, $\mathbf I$ is the unit tensor, $\mathbf g =(0,0,-g) $ is the gravitational acceleration, $\gamma$ is the specific heat ratio and the other symbols have their usual meaning. 
We assume the medium to be an ideal gas.
The $S$ terms are source terms in the equations to give the changes in density, momentum and energy of the system due to the addition of mass.
The method for mass addition is explained in Section \ref{ADD}.
It should be noted that the energy equation does not consider the energy balance properly, as thermal conduction and radiative heating and cooling are neglected.
These terms are important for formation of prominences by creating evaporation and condensation of plasma \citep[e.g.,][]{XIA2012, LUNA2012}, but for the purposes of understanding the response of the magnetic field to high density material, the equation we use is sufficient for our purposes.

A two-step Lax--Wendroff scheme based on the scheme presented in \citet{UGAI08} is used.
We take $\gamma=1.6$.
Damping terms are applied to the momentum terms to allow a relaxation to a new equilibrium.
The simulations are carried out on a $500 \times 500$ grid with uniform grid spacing of $dx=dz=0.01\Lambda$

\subsection{Initial Setting and Boundary Conditions for the Models}

\begin{figure*}[ht]
  \begin{center}
\includegraphics[width=7.5cm]{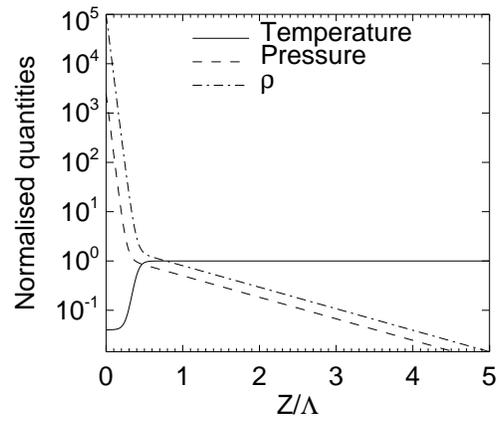}
  \end{center}
  \caption{Initial distribution of the hydrodynamic variables.}
\label{model_hydro}
\end{figure*}

The model applied here is an idealised solar atmosphere with hot $1$\,MK isothermal corona above a $40,000$K isothermal photosphere that extends for 10 photosphere pressure scale heights ($0.4\Lambda$) from the bottom of the calculation domain (see Figure \ref{model_hydro}).
Using this model the equations are non-dimentionalized using the sound speed ($C_s=10^7$\,cm\,s$^{-1}$), the pressure scale height ($\Lambda=C_s/(\gamma g)= R_gT/(\mu g)= 5.5 \times 10^9$\,cm) and take the density at the base of the corona above the transition region ($\rho_N =10^{-15}$\,g\,cm$^{-3})$.
We define the characteristic timescale ($\tau$) as the sound crossing time over one coronal pressure scale height giving $\tau=\Lambda/C_s=550$\,s.

The plasma $\beta$ for both models is defined at the respective o-point.
As each model has a different ratio of poloidal to axial field strength, the plasma $\beta$ is set so that the $B_y$ (axial) component of the field is approximately equivalent in each model.
To give a quantification of the strength of the horizontal ($B_x$) component of the magnetic field, for the region between the transition region and the o-point (x-point and o-point for Model 2), the ratio $|Max(B_y)/Max(B_x)|$ is $\sim 2.0$ for model 1 and $\sim 11.2$ for model 2.

The boundary conditions used are the same as those for the NLFFF calculation for the top boundary and the two side boundaries, i.e. a symmetric boundary which cannot be penetrated by the magnetic field.
The bottom boundary used is a symmetric boundary that can be penetrated by the magnetic field to reduce numerical issues in this region.
The magnetic field is changed by altering the vector potential over the bottom five grid points.
To remove the Lorentz force that this would create, a forcing term is added that is the equal and opposite of this force.
Also, this region is in a dense, high $\beta$ regime, therefore it cannot significantly affect the formation of the new equilibria.

\subsection{Method for Addition of Mass}\label{ADD}

Mass is added with a characteristic timescale of $t=2\tau$, i.e. in the region where mass is being added the density of each grid point should increase by one times the coronal mass in $t=2\tau$.
In this study we will look at density increases ($\rho'$) of factors $10$ and $25$ from the coronal density, therefore the input time for the mass addition is $t_{input}=20\tau$ and $50\tau$ respectively, after this time the source term $S_{\rho}$ in Equation \ref{cont} is set to $0$.
To define the height range over which the mass is added we use the distance between the flux tube o-point (${H_2}^n$) and the highest of either the height of the transition region or the x-point (${H_1}^n$), where the distance between these two points is defined as $\Delta H^n$ (the superscript $n$ shows the values at the $n$th time step).
Mass is input over the height range $\in H_1+ [0.1\Delta H,0.8\Delta H]$.
The characteristic width associated with the mass addition is $W_p=0.16\Lambda$ for most cases presented in this paper. 
The equation defining the source term for the mass addition for time step $n+1$ ($S _{\rho}(t^{n+1}) $) is as follows:
\begin{align}\label{MASSADD}
S_{\rho}(t^{n+1})  = & \frac{0.1}{\tau}\frac{ \rho_N}{4 \cosh(x/0.5W_p)}\left[\tanh\left(\frac{z-{H_{1}}^n+0.8 \Delta H^n}{0.07 \Delta H^n}\right)+1\right]  \\
{} & \times \left[1 -\tanh\left(\frac{z-{H_{1}}^n+0.1 \Delta H^n}{0.07 \Delta H^n}\right)\right] \notag
\end{align}

Once the new mass has been added at each new time step $n+1$ (during the addition of mass period), the conservative variables are recalculated.
This means that the addition of mass changes the momentum and energy of the system.
The addition of mass is not performed in a way that the temperature stays constant, as we aim to create cool dense regions in the corona.

\subsection{Conditions for Determining an Equilibrium}

With the MHD relaxation, conditions need to be imposed to determine when an equilibrium has been reached and when the time-marching scheme can be stopped.
To determine that an equilibrium has been reached (in fact it is a state that is approximately an equilibrium, but with small finite velocity) a maximum condition for the kinetic energy (KE) is used.
For any time $t$ such that $t > t_{input}$, if MAX(KE)/SE $< 5 \times 10^{-5}$ then the relaxation is determined to have reached equilibrium.
Here SE denotes the initial energy at the o-point given by SE$=p(o-point)/(\gamma-1)+B(o-point)^2/8\pi$.
An upper time limit for the relaxation was also imposed as $MAX(t_{relax})=t_{stop}(\beta)$, where $t_{stop}(\beta=0.4)=200\tau$, $t_{stop}(\beta=0.1)=150\tau$, $t_{stop}(\beta=0.04)=100\tau$ and $t_{stop}(\beta=0.01)=100\tau$.
This difference in time is used to reflect the higher speed at which information can travel through the calculation domain in the cases with lower plasma $\beta$.
If the condition MAX(KE)/SE $< 5 \times 10^{-5}$ is not satisfied during this time period, it is determined that no equilibrium is reached.

\section{Prominence Equilibrium}\label{IDEAL}

\subsection{Equilibrium of Model 1}\label{Model1}

Here the magneto-hydrostatic equilibrium that results from the addition of mass to an initially force-free magnetic field in a model solar atmosphere is described.
The mass was added to the system following the method described in Section \ref{ADD}.
The atmosphere was then allowed to relax to a new equilibrium and this equilibrium was investigated.
Six different cases were used in this investigation.
These are case a ($\beta=0.4$ and $\rho'=10$), case b ($\beta=0.4$ and $\rho'=25$), case c ($\beta=0.1$ and $\rho'=10$), case d ($\beta=0.1$ and $\rho'=25$), case e ($\beta=0.04$ and $\rho'=10$) and case f ($\beta=0.04$ and $\rho'=25$).
In this subsection and in Section \ref{COMPLEX}, for all figures, the cases are marked with their appropriate letter.

\begin{figure*}[ht]
  \begin{center}
\includegraphics[width=11cm]{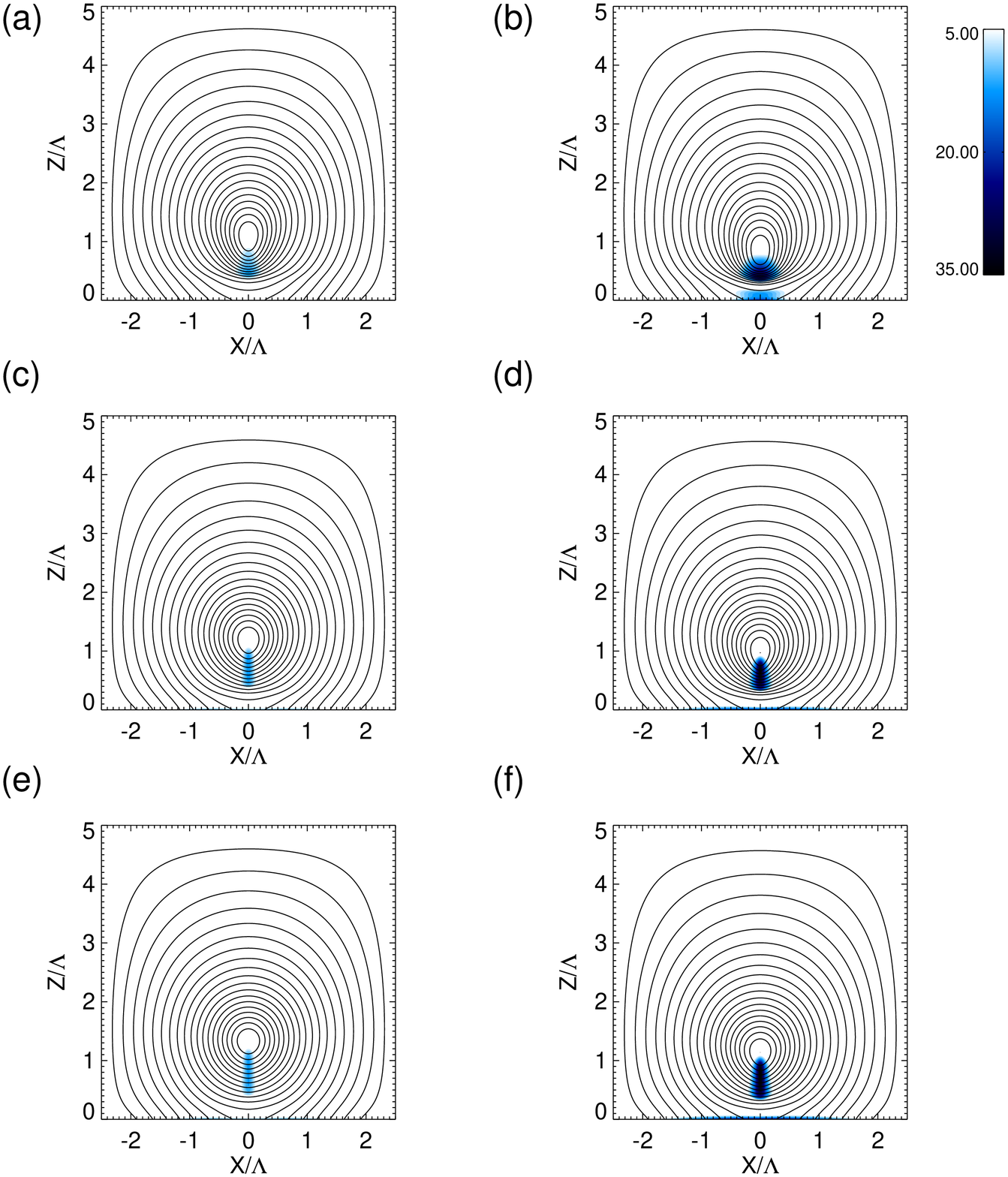}
  \end{center}
  \caption{Change in density distribution and magnetic field distribution (contours show magnetic vector potential) for model 1 in the six different cases. Going from left to right shows the increase in mass added and from top to bottom shows the decrease in plasma $\beta$. The global magnetic arcade does not show any significant change, but there is a drop in the height of the o-point and a stretching of the magnetic field of the prominence.}
\label{Model1_dist}
\end{figure*}

Figure \ref{Model1_dist} shows the magnetic field distribution and the change in density from the initial distribution ($\rho'$) for the six different parameter sets.
A few trends are obvious from this figure.
The width of the prominence becomes larger and the height smaller for a greater density, the same trend can be seen for the plasma $\beta$.
For the overlying field lines, there is no great difference irrespective of the plasma $\beta$ or $\rho'$.
However, there are noticeable changes to the magnetic field structure for the region of the magnetic field that supports the prominence material.
The height of the o-point falls, with a greater fall for a larger $\rho'$ and $\beta$, as well stretching of the magnetic field below the o-point and the compression of the magnetic field near the base of the prominence.
It should be noted that the density increases that are visible at the bottom of the simulation domain, though comparable to the density increase in the prominence, only represents an increase of approximately $0.01$\,\% of the surrounding density in the photosphere.

\begin{figure*}[ht]
  \begin{center}
\includegraphics[width=11cm]{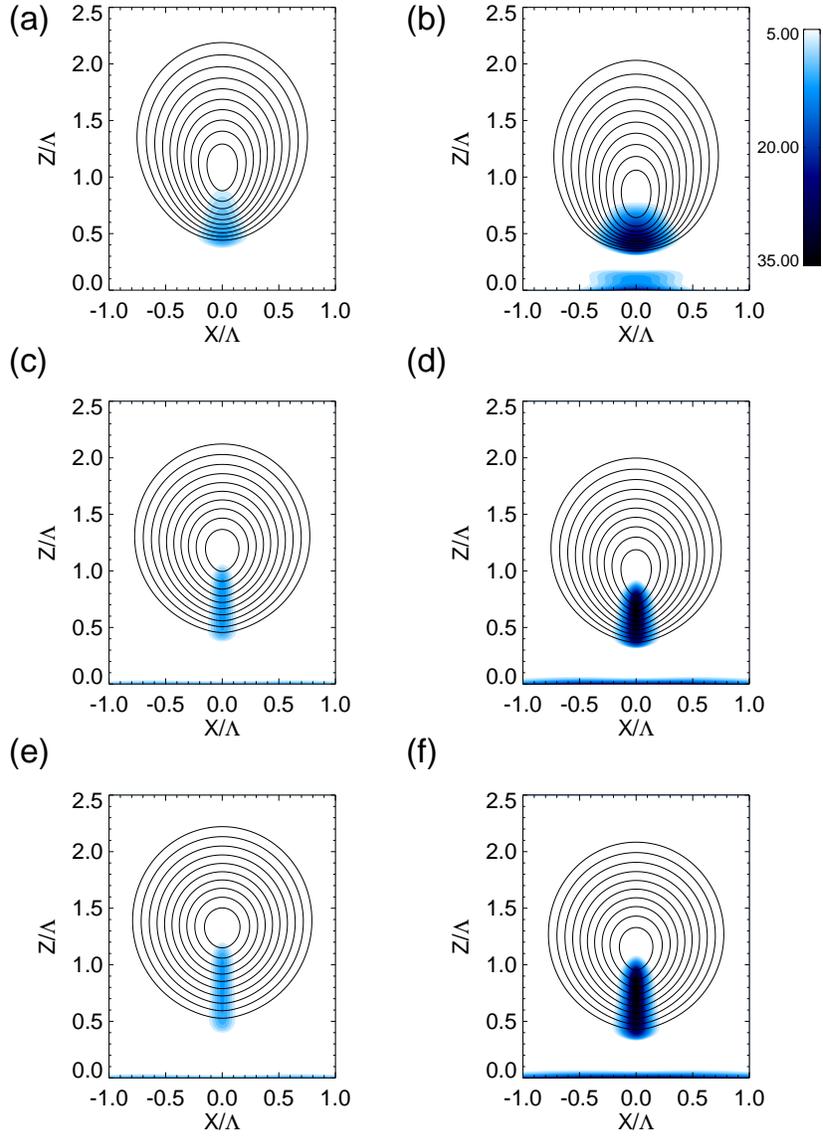}

  \end{center}
  \caption{Local change in density distribution and magnetic field distribution (contours show magnetic vector potential) for model 1 in the six different cases looking only at the prominence region. Going from left to right shows the increase in mass added and from top to bottom shows the decrease in plasma $\beta$. The drop in the height of the o-point and stretching of the magnetic field are clearly visible, especially for the high $\beta$ cases.}
\label{Model1_local}
\end{figure*}

\begin{figure*}[ht]
  \begin{center}
\includegraphics[width=10cm]{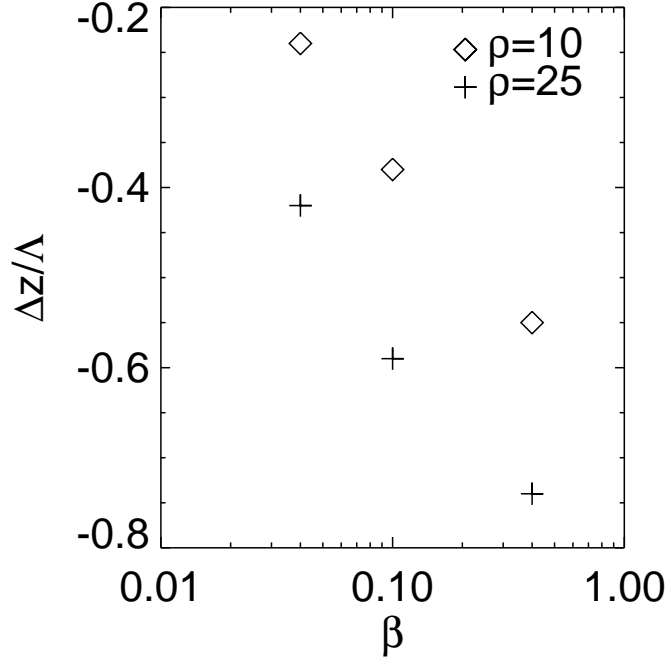}

  \end{center}
  \caption{Change in height of the o-point in relation to the plasma $\beta$.}
\label{Model1_op}
\end{figure*}

An expanded view of the prominence density distribution and the magnetic field of the prominence is shown in Figure \ref{Model1_local}.
This clearly shows the structure described previously.
As with the global field, the field lines plotted in the figure that are furthest from the o-point do not change greatly, whereas those field lines that are closer to the o-point undergo more stretching to support the prominence material.
The compression of the magnetic field can be seen.
Case f, with the $\rho'=25$ and $\beta=0.04$ has a structure that represents that of the expected structure of the quiescent prominence that is supported by magnetic tension.
The drop in o-point with height is given in Figure \ref{Model1_op}.
The change in height from the original o-point is plotted against the plasma $\beta$.
It can be seen that the change in height is strongly related to the change in plasma $\beta$.

\begin{figure*}[ht]
  \begin{center}
\includegraphics[width=12cm]{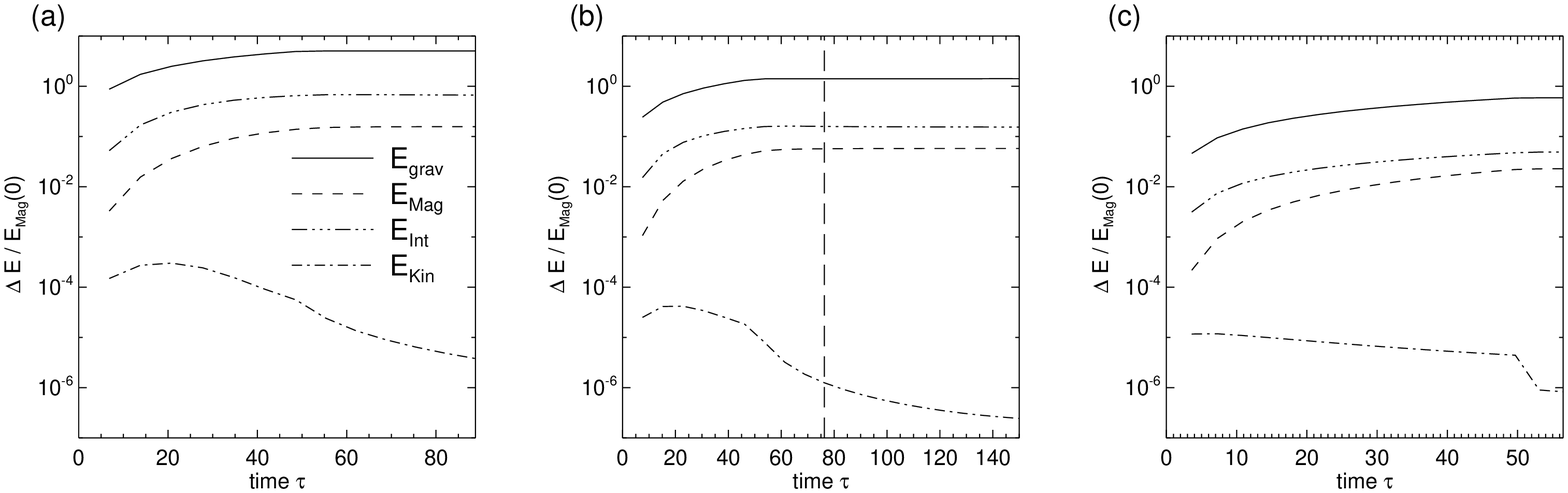}
  \end{center}
  \caption{Temporal evolution of the change in energy in the simulation domain for cases b (panel (a)), d (panel (b)) and f (panel (c)) of model 1.}
\label{Model1_en}
\end{figure*}

Figure \ref{Model1_en} shows the evolution of the magnetic, kinetic, gravitational potential and internal energies with time for cases b, d and f.
The plots in each simulation show the change in energy from the initial state, with the values shown are normalised by the initial magnetic energy (IME) of the respective simulation.
These figures show that after the mass has been added to the flux rope, the system relaxes with a steady decrease in total kinetic energy (TKE) until the TKE/IME $<10^{-6}$ and the other energies have reached constant values.
The vertical dashed line in panel (b) shows the approximate time at which the conditions for equilibrium are satisfied, with the continued evolution shown for reference.

\begin{figure*}[ht]
  \begin{center}
\includegraphics[width=14cm]{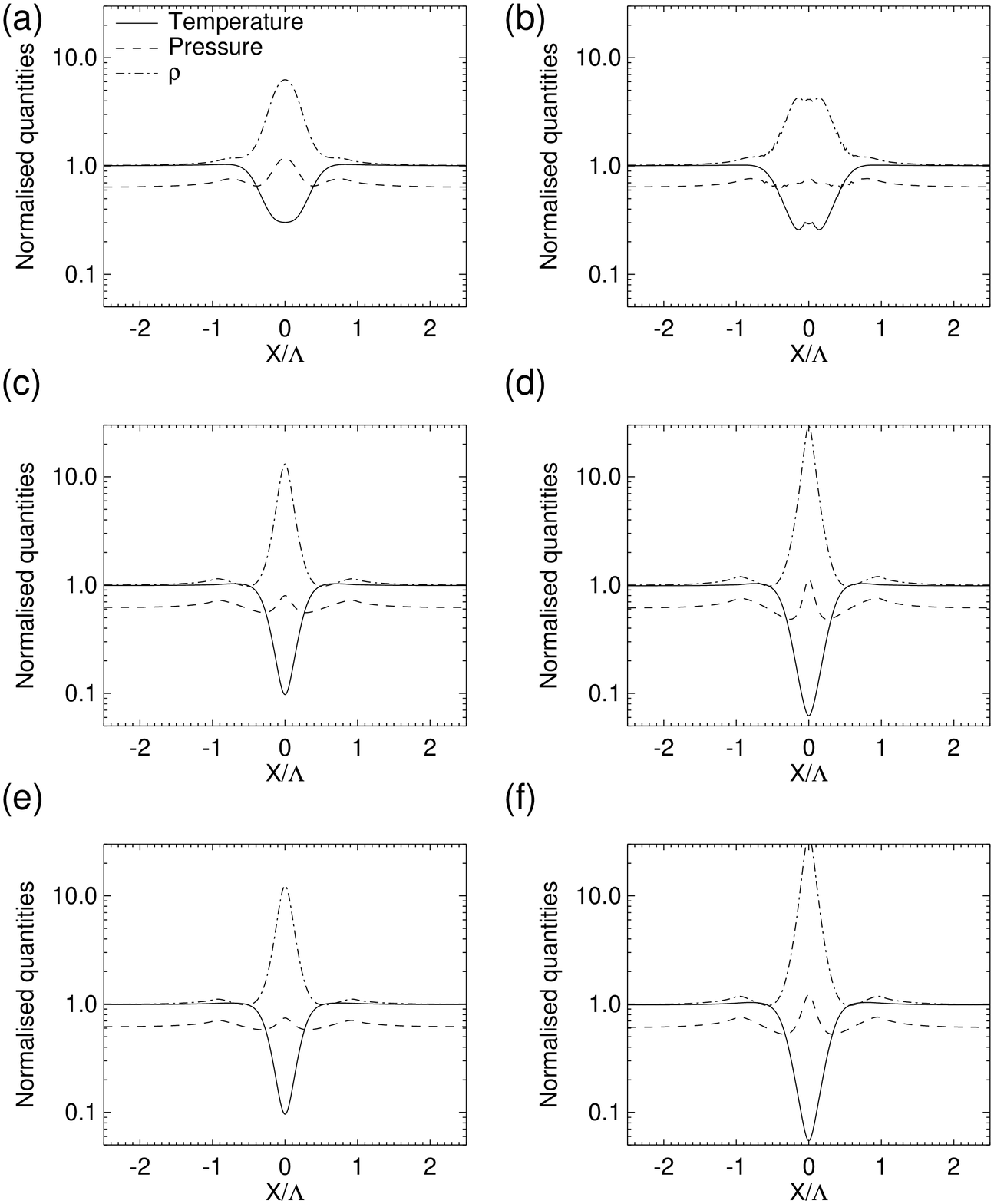}

  \end{center}
  \caption{Horizontal distribution of the hydrodynamic variables for the six different parameter sets for model 1. The distribution is taken at the height $z = 0.8 \Lambda$.}
\label{Model1_hor}
\end{figure*}

Next, the horizontal and vertical distribution of the hydrodynamic variables are given.
The horizontal distributions at $z=0.8\Lambda$ are shown in Figure \ref{Model1_hor}.
For the models where a prominence is supported by magnetic tension show a thin, high density prominence region, the horizontal pressure distribution shows that the pressure at the centre of the prominence has increased by a factor of two when compared to the background pressure.

\begin{figure*}[ht]
  \begin{center}
\includegraphics[width=14cm]{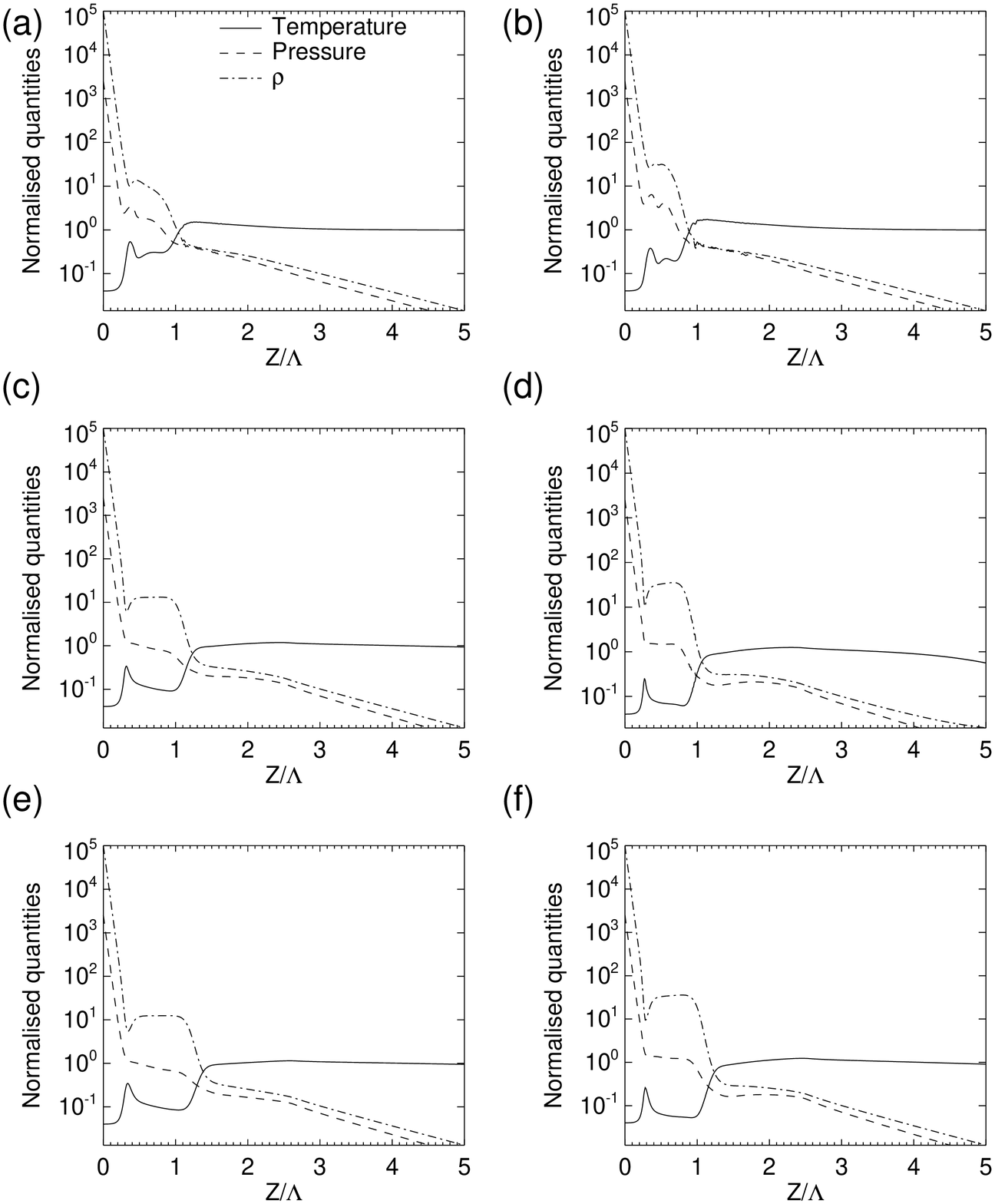}

  \end{center}
  \caption{Vertical distribution of temperature, pressure and density for the six parameter sets of model 1. The distributions are taken at the horizontal position $x=0\Lambda$.}
\label{Model1_ver}
\end{figure*}

Figure \ref{Model1_ver} gives the vertical distributions of the hydrodynamic variables at $x=0$.
The two extreme cases, which imply two different mechanisms of support, are shown in panels (b) and (f). 
The density distributions shown are very different.
The distribution shown in panel (b) decreases exponentially, implying that the material is supported by a pressure gradient (either gas or magnetic).
The distribution shown in panel (f) has a density distribution that is almost constant with height in the prominence, as well as a pressure distribution that is also almost constant with height.
In this case the dense material is supported by magnetic tension.

\begin{figure*}[ht]
  \begin{center}
\includegraphics[width=14cm]{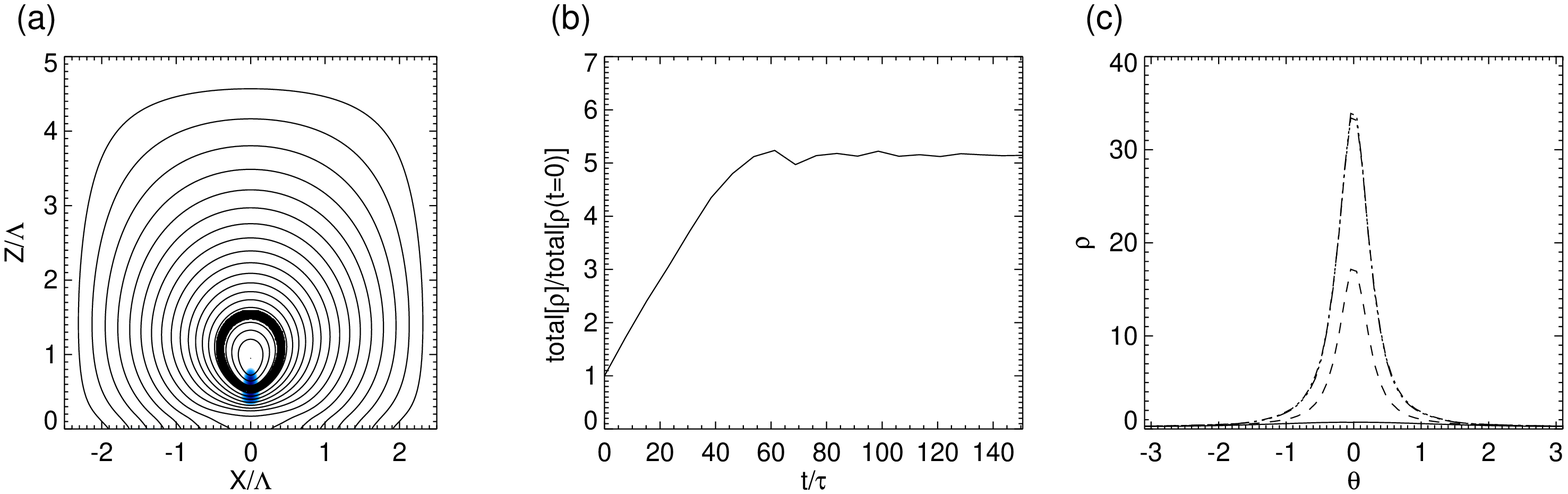}

  \end{center}
  \caption{Figure showing the evolution of the total mass of a ring section of the flux rope. Panel (a) shows the flux tube with the prominence, with the thick black line showing the region where the mass evolution is calculated. Panel (b) shows the evolution of mass with time. Panel (c) shows the density by position (as angle $\theta$ around the o-point) at time $t=0.0$, $30.7$, $61.3$ and $91.2$.}
\label{Model1_masscon}
\end{figure*}

Figure \ref{Model1_masscon} shows the mass conservation in case d (shown in Figure \ref{Model1_dist}).
The thick black line in panel (a) shows the region in which the mass conservation is calculated.
Panel (b) gives the total mass (normalised by the initial value) in the black region.
The mass increases over the period where mass is added, then the system relaxes to the equilibrium.
The total mass loss from the region (through numerical mass diffusion) at the end of the calculation is approximately $0.5$\,\% of the total mass in the black region at the time that the mass injection is halted.
Panel (c) gives the average density in the black region in terms of angle around the o-point of the flux tube at times $t=0.0$, $30.7$, $61.3$ and $91.2$.

\begin{figure*}[ht]
  \begin{center}
\includegraphics[width=14cm]{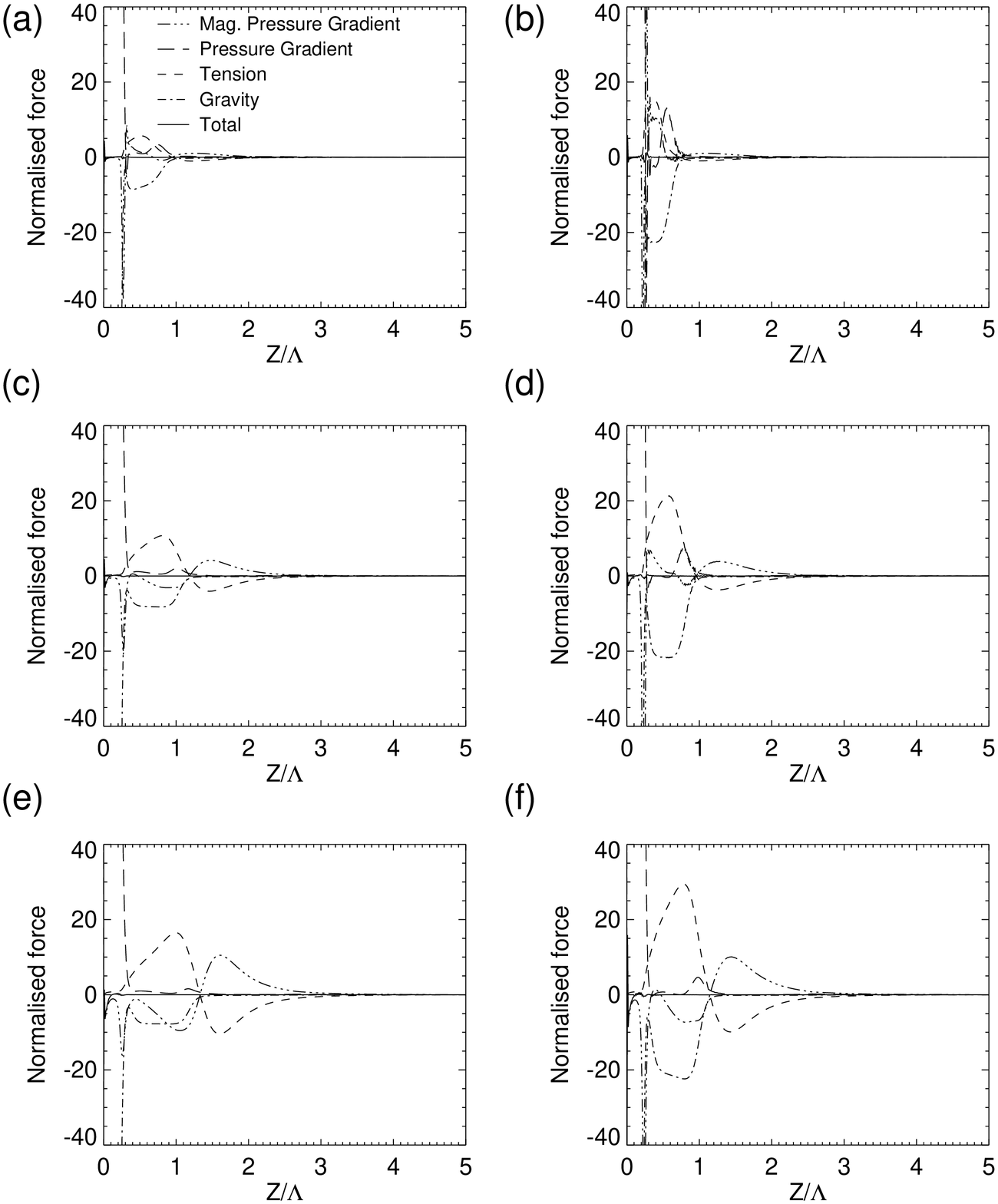}

  \end{center}
  \caption{Force distribution for the six different parameter sets for model 1. The distributions are taken at the horizontal position $x=0\Lambda$.}
\label{Model1_force}
\end{figure*}

Figure \ref{Model1_force} shows the force distributions in the $z$-direction at $x=0$ for the final equilibriums of the different parameter sets for model 1.
The distributions shown in panels (c)--(f) represent the distribution that would be expected for a prominence where the material is supported by magnetic tension.
That is to say the gravitational force and downward acting magnetic pressure gradient (that is part of the initial distribution of the force-free magnetic field) are balanced by the upward acting magnetic tension force.
The gas pressure gradient is not significant, apart from at the top of the prominence as a result of the increased prominence gas pressure dropping off to match with the lower coronal gas pressure.

The panels (a) and (b) show a very different distribution of the forces to the parameter sets discussed in the previous paragraph. 
In these parameter sets, there is no longer a downward acting magnetic pressure gradient.
In panel (d), which appears to be a borderline case, this gradient is approximately $0$, but for panels (a) and (b) the direction in which the magnetic pressure gradient is working has changed.
Therefore, in these two cases it can be said that the prominence material is supported by a combination of the magnetic pressure and magnetic tension forces as well as a contribution from the gas pressure.
When looking at these cases in Figure \ref{Model1_local}, it can be seen that these equilibria can be viewed as failed prominences.
This implies that, as expected, for prominences formed in dips of flux tubes, magnetic tension is the force that supports the prominence material.

Figure \ref{Model1_beta} shows the vertical distribution of plasma $\beta$ at $x=0$.
The plasma $\beta$ is calculated including all the components of the magnetic field (solid line) and also only including the $B_x$ component (dashed line).
Looking at panels (b), (d) and (f) shows a change in the plasma $\beta$ distribution in the prominence material ($\sim 0.5$\,--\,$1\Lambda$).
Panel (b) shows a peak in the plasma $\beta$, panel (d) shows that the plasma $\beta$ is almost constant and panel (f) shows a plasma $\beta$ that decreases with height, implying that the dominant axial field still exists.
Looking only at the $B_x$ component, which is important for the tension support, shows that it takes approximately a plasma $\beta \sim 0.8$ in the prominence (see panel (c)) to support $\rho'=10$ and approximately a plasma $\beta \sim 0.3$ in the prominence (see panel (f)) to support $\rho'=25$.

\subsection{Method of Mass Support}

To investigate the nature of the support of the prominence material by magnetic tension, we take the parameter set with $\beta =0.04$ and $\rho'=25$.
Figure \ref{Model1_mag_dist} give the vertical distributions of $B_x$ and $B_y$ at $x=0$ and the horizontal distribution of $B_z$ at $z=0.8$.
The vertical distributions show that the strength of the $B_x$ and $B_y$ components of the magnetic field have increased in strength by a factor of approximately $1.34$ and $2.98$ respectively at the height $z=0.8\Lambda$.
These increases in the field strength are a direct result of the compression of the magnetic field as a result of the increased density dragging down the magnetic field.
The $B_z$ distribution shows that the maximum magnetic field strength at the height $z=0.8\Lambda$ increases by a factor of $1.66$ and the gradient of the magnetic field is greatly increased.
This combination of the larger gradient in the vertical field and the compression of the horizontal field allow the coronal magnetic field to support the prominence material.

Figure \ref{Model1_mag_lines} shows the magnetic field lines for the initial magnetic field and the final prominence equilibrium.
It can be seen from this figure that the curvature of the magnetic field at the base of the prominence is not significantly changed by the addition of mass, but that the magnetic field lines have accumulated.
The curvature of the magnetic field, however, increases with height, which follows the decrease in the horizontal component of the magnetic field.
This implies that in our model the support for the prominence material happens in the following way.
\begin{itemize}
\item The dense material falls due to gravity, pulling the magnetic field with it, stretching the magnetic field.

\item The stretched magnetic field exerts a tension force on the whole of the flux tube, which then is pulled downward.

\item As the flux tube drops, the magnetic field in the prominence compresses especially at the base of the prominence, meaning that a reduced curvature can produce a stronger tension force

\item This tension force, where the curvature of the magnetic field changes significantly with height, can now support the prominence.

\end{itemize}

\subsection{Varying Mass Input Width}

Here we investigate the change in the prominence structure that results from the change in the width over which mass is input.
The parameter set used to investigate this is $\beta=0.04$ and $\rho'=25$.
To look at the difference in width, the mass is added with widths of $W_p=0.16\Lambda$, $0.32\Lambda$ and $0.48\Lambda$ used in Equation (\ref{MASSADD}).

Figure \ref{Model1_wid} shows the global and local magnetic field and density distribution for, from top to bottom, widths of $W_p=0.16\Lambda$, $0.32\Lambda$ and $0.48\Lambda$.
It can be seen that the mass input with greater width results in a lower o-point and a higher central density.
This higher density comes because, even though the o-point drops, this does not significantly change the curvature of the magnetic field.
As there is a component of gravity that works along the direction of the magnetic field, the prominence will contract until the pressure gradient in the prominence is great enough to balance the gravitational force.
Therefore, for greater widths for adding the prominence material, the final prominence density becomes higher.

To analyse the width of the prominence and how it changes with the input width, the prominence density and pressure distributions at the height $z=0.6\Lambda$ is fitted with a Gaussian distribution and the full width half maximum (FWHM) of this fitted Gaussian distribution is then taken as the width of the prominence.
For the three models, the FWHM of the fitted Gaussian distributions are $0.1\Lambda$, $0.18\Lambda$ and $0.26\Lambda$ for the density and $0.1\Lambda$, $0.20\Lambda$ and $0.26\Lambda$ for the pressure.
Therefore the width of the prominence can be seen to have contracted by a factor of $1.5$\,--\,$2$.
This implies that, even with the drop in o-point height, which can reduce the curvature of the magnetic field, the accretion of mass through contraction of the prominence along the direction of the magnetic field is sufficient to pull in the prominence material.

Looking at the vertical force balance, see Figure \ref{Model1_wid_force}, it is clear that the input width of $W_p=0.48\Lambda$ is approaching being a failed prominence due to the larger support from pressure gradients.
Therefore, it is unlikely that $\beta=0.04$ is sufficient to support prominences with higher mass.

\subsection{Investigation of More Complex Magnetic Field Model}\label{COMPLEX}

Here we will present the results from the mass addition for a different magnetic field model shown in Figure \ref{model} (b).
This model is called model 2.
This model differs from the model used for the previous results in three key areas:
\begin{itemize}
\item The o-point is initially higher by approximately $0.5\Lambda$
\item The ratio of $B_y$ to $B_x$ is $\sim 11$
\item There is an x-point below the o-point
\end{itemize}
With these differences, there is a significant change in how the system behaves.
In fact, for the simulated parameters, the model does not reach an equilibrium.
It should be noted here that the plasma $\beta$ values investigated here are $\beta=0.1$, $0.04$ and $0.01$ to give a stronger $B_x$ component of the magnetic field to increase the ability of the field to support the plasma.

The reason for the significant change from the previous model can be seen in Figure \ref{Model2_local} which shows the local distribution of the magnetic field and the density change.
As with the previous model, for the global field the overlying arcade remains reasonably unchanged, but it can be seen in Figure \ref{Model2_local} that a long, vertical current sheet develops in the prominence magnetic field and with reconnection occurring at the x-point.
The change in topology that this implies, means that there is dissipation of the horizontal field component that is needed to support the prominence material.
Therefore, while reconnection is occurring it is not possible for the prominence material to form a steady state.
It can also be seen that for the $\beta=0.1$ simulations, a long thin current that forms as the magnetic field is stretched by the falling plasma. 

The temporal evolution of the energies for cases b (panel (a)), d (panel (b)) and f (panel (c)) is displayed in Figure \ref{Model2_en}.
In all three cases, the maximum KE has not fallen below the set threshold and for case b it can be seen that the TKE is increasing as the simulation progresses.
The reason why the maximum KE does not fall below the set criteria is presented in Figure \ref{Model2_vervel}, which shows the temporal evolution of the vertical velocity at $x=0$ for case f.
Peaks in the velocity can be found at the position of the x-point, highlighting this region's importance for the continued evolution of the system.

The relation between the drop in the o-point and how distended the magnetic field becomes for this model is very different from the previous model.
This is because, as mentioned previously, a long, vertical current sheet develops above the prominence where only a small component of the horizontal field remains.
Where, in the previous case, the o-point would fall with the material, here the o-point is stretched into a line where there is no $B_x$ component of the magnetic field (this is discussed later).
In this respect, it can be seen that the movement of the o-point is determined by the amount of tension that is needed by the magnetic field to support the mass and the strength of the magnetic pressure at the o-point.
The change in the height of the o-point is shown in Figure \ref{Model2_opoint}.

The vertical distribution (at $x=0\Lambda$) is shown in Figure \ref{Model2_vertical}.
This shows, apart from panel (e), similar distributions of the prominence density to the failed prominences shown in Section \ref{Model1}.
Only in panel (e) does the dense material appear to be supported to some extent by magnetic tension.
This is not surprising given the significantly weaker horizontal magnetic field distribution.

Another point of note is the drop of density between the prominence and the region below.
This is a result of the x-point, which creates a thin region with coronal density beneath the prominence, this can be seen in Figure \ref{Model2_local}.
This region may be unstable to the magnetic Rayleigh--Taylor instability, as observed in prominences by \citet{BERG2008,BERG2010} and studied numerically by \citet{HILL2011b,HILL2012a}.

It should be noted again that this model does not reach an equilibrium, as can be seen in Figure \ref{Model2_force}.
The main difference between this model and the previous model shown in Section \ref{Model1} is that the main upward oriented force in the prominence region is gas pressure (i.e. see panel (b) of Figure \ref{Model2_force}).
Therefore, these prominences are only using the magnetic field to collimate the material and are using gas pressure to support the material.
This implies that the field strength of the horizontal field is not sufficient and so lower plasma $\beta$ values are necessary for this 2.5D model, where this tendency can be seen in the Figure. 

Figure \ref{Model2_beta} shows the plasma $\beta$ for this model..
Again both the plasma $\beta$ calculated using all the magnetic field components (solid line) and only the $B_x$ component are shown.
Though the plasma $\beta$ values are very small, looking at the $B_x$ component shows that the horizontal field is not strong enough to support the material through magnetic tension.
Only in the case presented in panel (e) has the horizontal field become close to strong enough to support the added mass.

Figure \ref{Model2_mag_dist} gives the vertical distributions of the $B_x$ and $B_y$ components of the magnetic field at $x=0$ and the horizontal distribution of $B_z$ at $z=1\Lambda$.
The distribution of $B_x$ (panel (a)) shows, as mentioned previously, that the o-point is no longer a point, but a vertical line with little or no $B_x$ component of the magnetic field.
As is known from the studies of the tearing instability \citep{FKR1962}, when this line reaches a critical length, the system is unstable to the formation of magnetic islands.
The distribution of $B_z$ which results in the current sheet in this $|B_x|\ll 1$ region is shown in panel (c).
This will result in reconnection which separates the prominence magnetic field into a separate flux tube embedded in the initial coronal flux tube.
Above the prominence region, the $B_y$ distribution of the magnetic field does not show any significant change, but there is an accumulation of the axial field at the base of the prominence.

Comparing the results from this case with the results from case 1, it is clear that there is a difference in the way the magnetic field evolves toward a magneto-hydrostatic equilibrium.
For the case presented in this subsection, the o-point is stretched resulting in the bottom of this region giving the large drop in the o-point in some cases shown in Figure \ref{Model2_opoint}, this stretching of the o-point into a line over which there is little to no $B_x$ component of the magnetic field, especially for the parameter case b as this has a weaker magnetic field, gives a long, thin current sheet.
The step like current sheet can be clearly seen for the $B_z$ distribution in this parameter set.
For parameter set f, where the magnetic field is stronger, this is not so pronounced, but the weak horizontal field is still noticeable.
This stretching of the o-point results from a divergence of the velocity field in the $z$-direction, which has its peak at the o-point.
This divergence would advect the horizontal field away from the o-point, creating a long, thin region where the horizontal field is either $0$ or close to zero (see Figure \ref{Model2_mag_dist} panel (a)).

For model 1, two different support mechanisms were found: pressure support and tension support.
For model 2 neither is found to be effective.
The tension support would not be able to work because the horizontal field is insufficiently strong to support the magnetic field through magnetic tension.
There is, however, a very large $B_y$ component, especially in the $\beta=0.01$ model, so it could be expected that pressure support may be possible.
This mechanism does not work in this model because of the reconnection at the x-point, which means that plasma trapped inside dips and supported by magnetic pressure, can escape from the system.
Figure \ref{den_loss} shows the change in the mass in the flux tube (normalised so the initial mass is 1).
As reconnection is occurring, it should be noted that the size of the fluxtube is decreasing in time.
From the peak in the mass, $28$\,\% of the mass is lost over a time of $\sim 100\tau$, over the same period only a maximum of $2$\,\% of the total mass loss can be attributed to cross field diffusion of mass out of the flux rope due to numerical diffusion.
Therefore, the reconnection allows the mass to escape the prominence.

\section{Summary and Discussion}\label{SUM}

In this study, we have looked at the support of mass against gravity by the Lorentz force of a coronal flux rope.
It was found that the case where the support of the mass has a large contribution from pressure gradients, the structure of the formed prominence was very different from observed prominences, therefore this case can be seen as a failed prominence.
However, when magnetic tension is the main force for the support of the dense material, then a prominence-like structure is formed.

The mechanism for the tension support results from two key processes.
The upper region of the prominence is supported by the stretching of the magnetic field, which results in higher tension.
However, at the base of the prominence, the support comes from the curvature of compressed magnetic field.
As the magnetic field is compressed, it does not need a strong radius of curvature to produce the same tension as the magnetic field at the top of the prominence.
The compression of the magnetic field is a direct result of the added mass, which pulls down the o-point so magnetic field accumulates at the base of the prominence.

Comparison between the two models showed that the drop in the o-point, and how much the magnetic field was stretched, depended heavily on the ratio of the axial field ($B_y$) to the horizontal field ($B_x$).
It appears that when this ratio is large, the height of the o-point becomes more stable and there is greater stretching of the magnetic field, whereas when the ratio is small then there is a greater drop in the o-point height with less stretching of the magnetic field.
However, due to the weak horizontal field and, in some part, to numerical reconnection in these models (at both the x-point and in the prominence current sheet), it has not been possible to achieve the magneto-hydrostatic equilibrium for the second model, therefore it is difficult to give these results with greater accuracy.

The reconnection found in model 2 was a result of the numerical dissipation of the magnetic field.
Therefore, it can be expected that the use of more grid points, or a numerical scheme that is less dissipative, would slow down the rate of the reconnection.
Assuming Sweet--Parker reconnection, where the reconnection rate is given by $v_{in}/v_{A}~S^{-1/2}$ with $v_{in}$ as the inflow velocity, $v_{A}$ as the Alfv\'{e}n velocity and $S$ as the Lundquist number, reduction in the numerical diffusion will slow down the rate at which magnetic field reconnects.
However, only in the theoretical limit of 0 error can it be expected that numerical reconnection would not occur allowing a true equilibrium to form.

\begin{figure*}[ht]
  \begin{center}
\includegraphics[width=14cm]{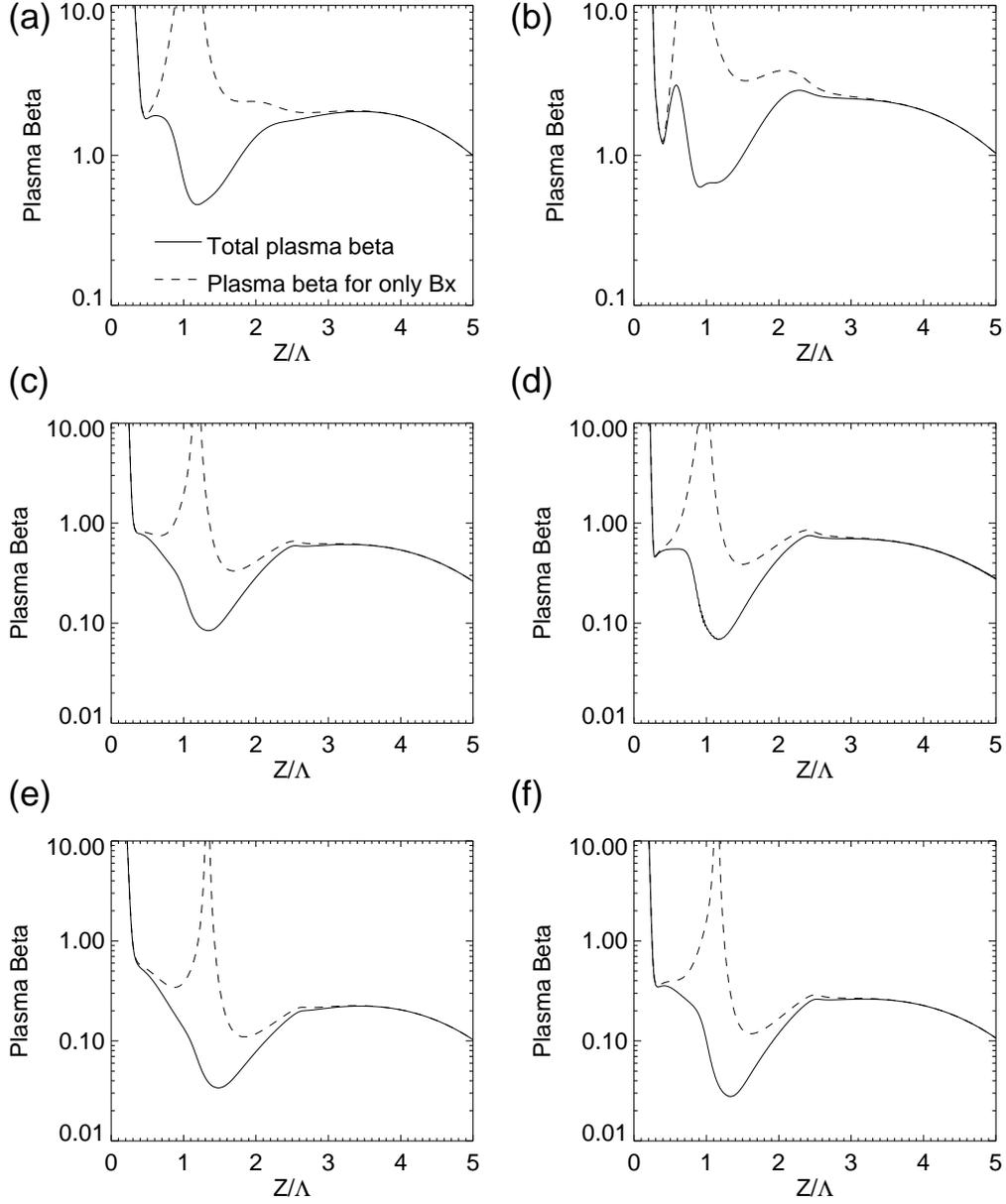}
  \end{center}
  \caption{Plasma $\beta$ distribution with height for the six different parameter sets for model 1 taken at $x=0$. The solid line shows the plasma $\beta$ for all magnetic field components, and the dashed line shows the plasma $\beta$ for the $B_x$ component, which is related to the amount of tension the magnetic field can exert.}
\label{Model1_beta}
\end{figure*}

\begin{figure*}[ht]
  \begin{center}
\includegraphics[width=16cm]{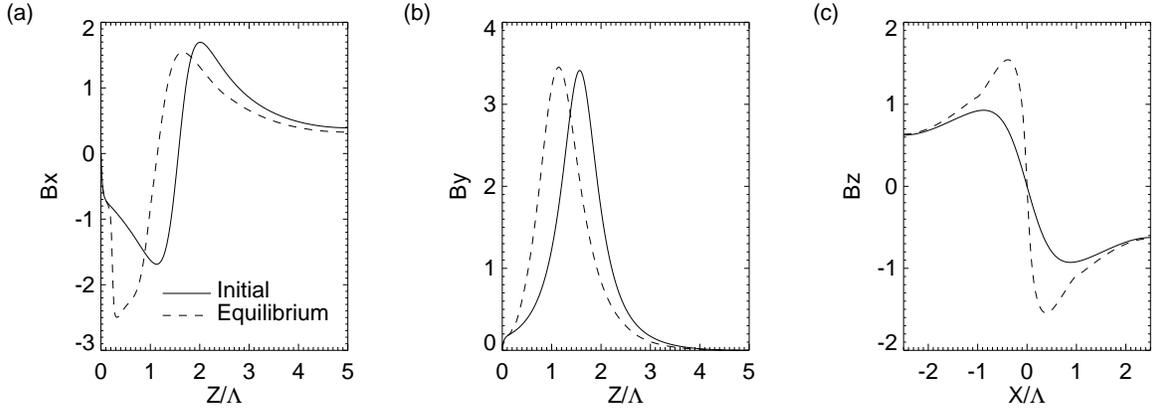}
  \end{center}
  \caption{Magnetic field distribution for model 1 parameter set f. The $B_x$ and $B_y$ distributions are taken at $x=0$ and the $B_z$ distribution is taken at $z=0.8\Lambda$.}
\label{Model1_mag_dist}
\end{figure*}

\begin{figure*}[ht]
  \begin{center}
\includegraphics[width=12cm]{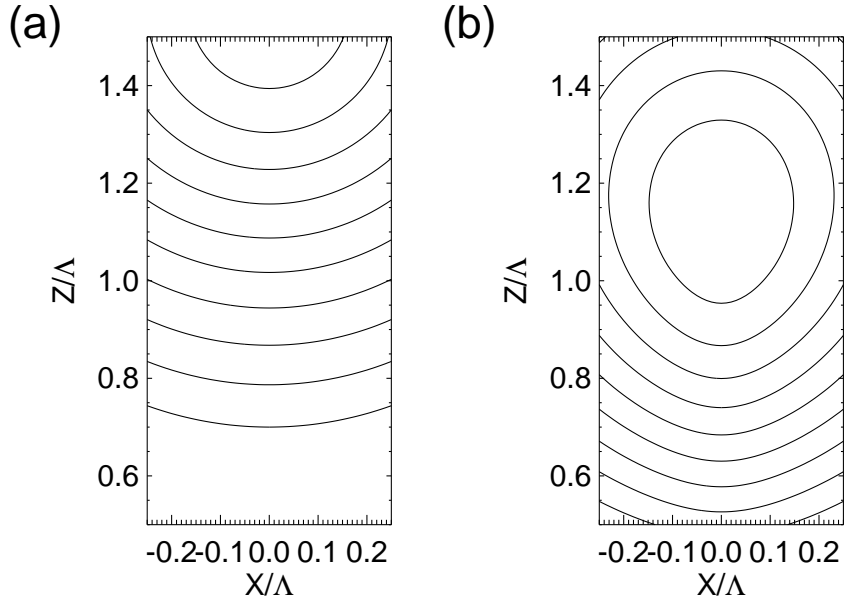}

  \end{center}
  \caption{Magnetic field distribution of the parameter set $\beta=0.04$ and $\rho'=25$ both (a) before the mass addition and (b) once the magneto-hydrostatic equilibrium has formed. Lines show the distribution of $A_y(x,z)$, where the same values of $A_y(x,z)$ are used to draw the contours for each plot.}
\label{Model1_mag_lines}
\end{figure*}

\begin{figure*}[ht]
  \begin{center}
\includegraphics[width=11cm]{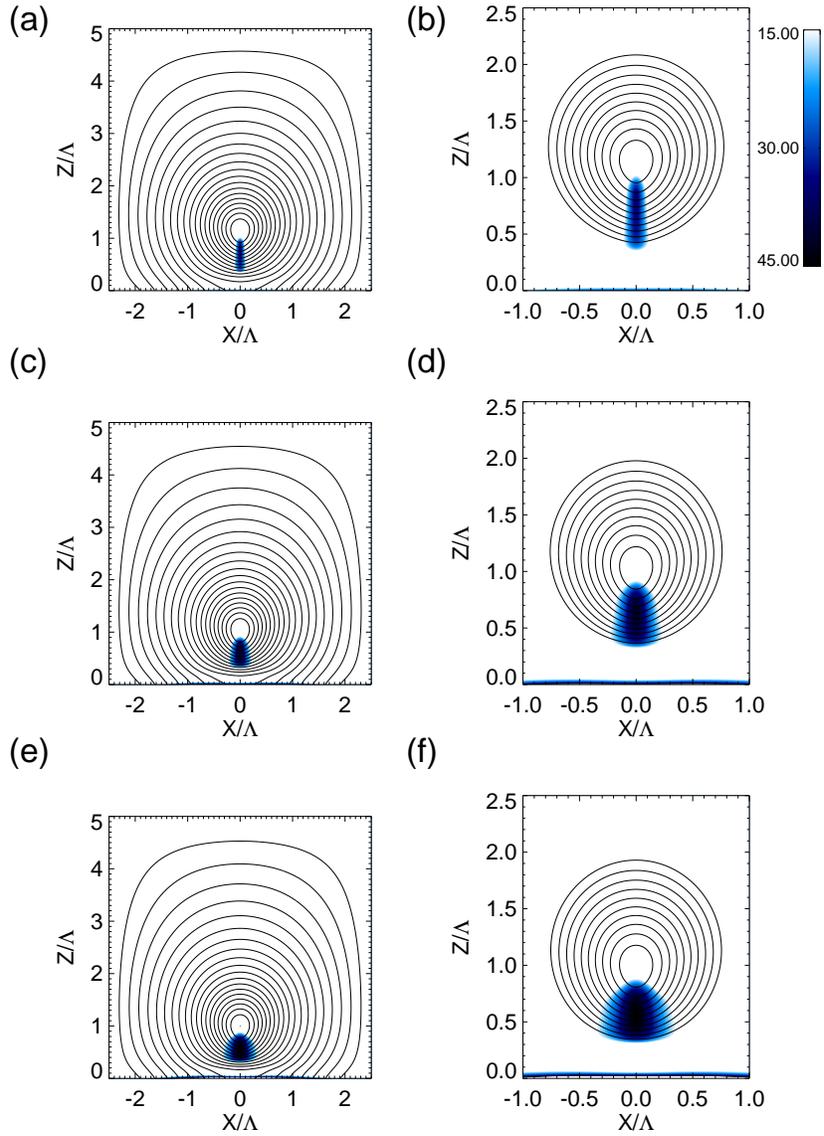}

  \end{center}
  \caption{Global and local change of the magnetic field and density distributions for the models with different input widths for the mass. }
\label{Model1_wid}
\end{figure*}

\begin{figure*}[ht]
  \begin{center}
\includegraphics[width=16cm]{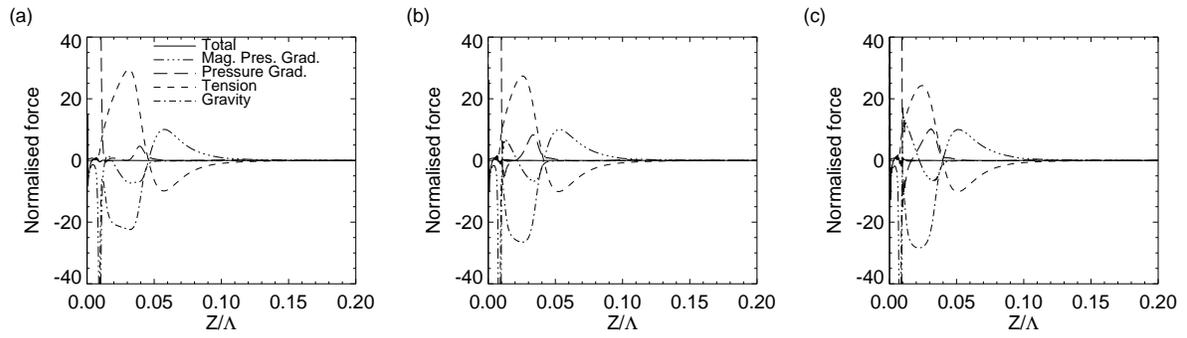}
  \end{center}
  \caption{Force distribution for the three different input widths for model 1. The distributions are taken at the horizontal position $x=0\Lambda$.}
\label{Model1_wid_force}
\end{figure*}

\begin{figure*}[ht]
  \begin{center}
\includegraphics[width=11cm]{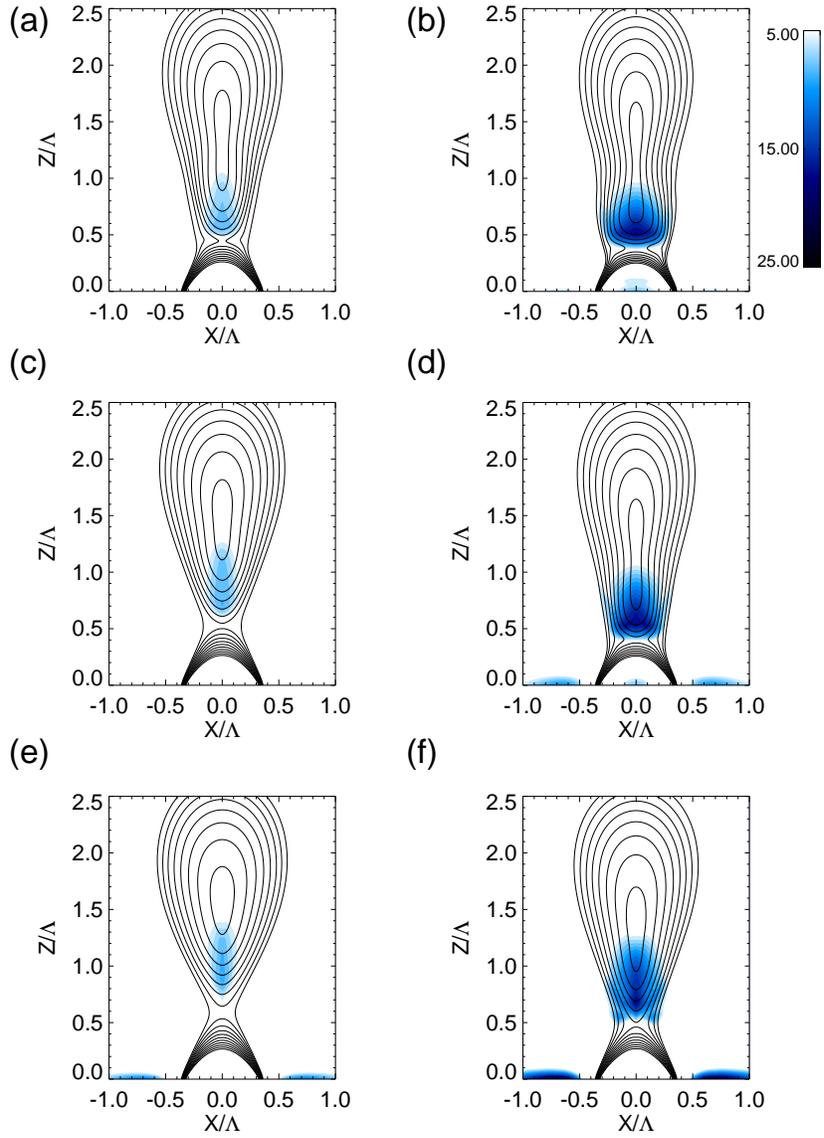}
  \end{center}
  \caption{Local change of the magnetic field and density distributions for model 2. }
\label{Model2_local}
\end{figure*}

\begin{figure}[ht]
 \centering
\includegraphics[width=12cm]{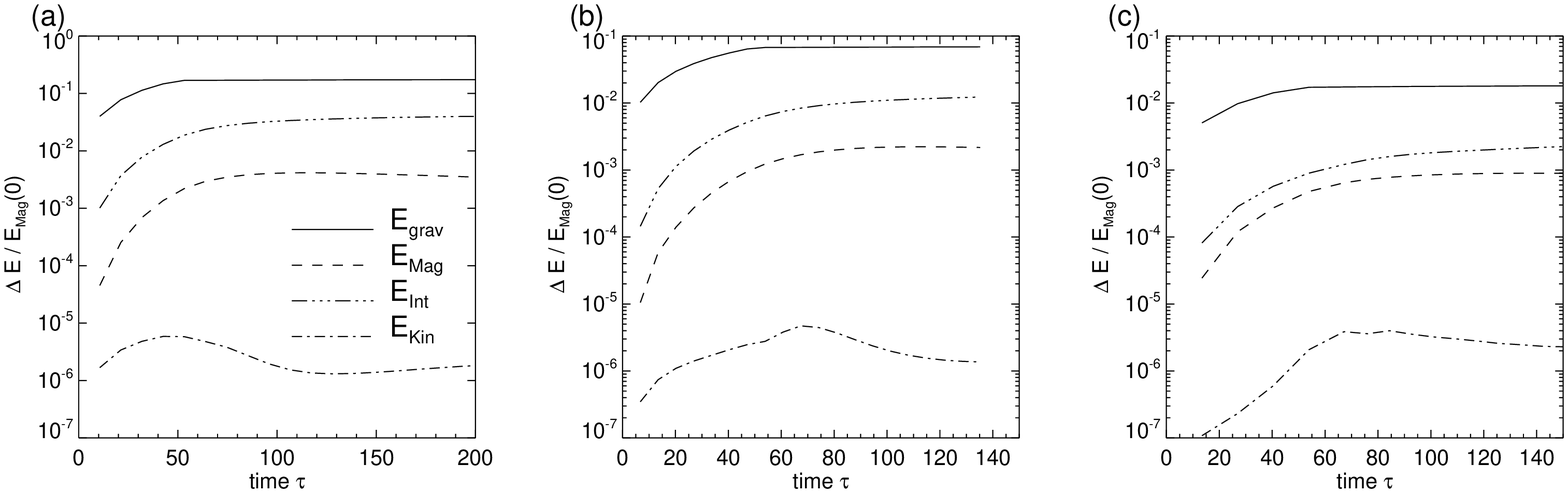}
  \caption{Temporal evolution of the change in energy in the simulation domain for cases b (panel (a)), d (panel (b)) and f (panel (c)) of model 2.}
\label{Model2_en}
\end{figure}

\begin{figure}[ht]
 \centering
\includegraphics[width=4cm]{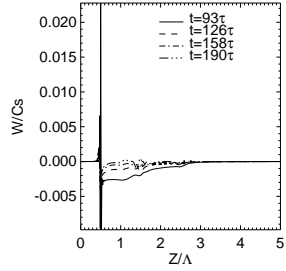}
  \caption{Temporal evolution of the vertical velocity ($w$) at $x=0\Lambda$. This highlights that even though the velocities in the prominence are small, a sharp jump in velocity (showing a converging flow) exists at the x-point.}
\label{Model2_vervel}
\end{figure}

\begin{figure}[ht]
\centering
\includegraphics[width=8cm]{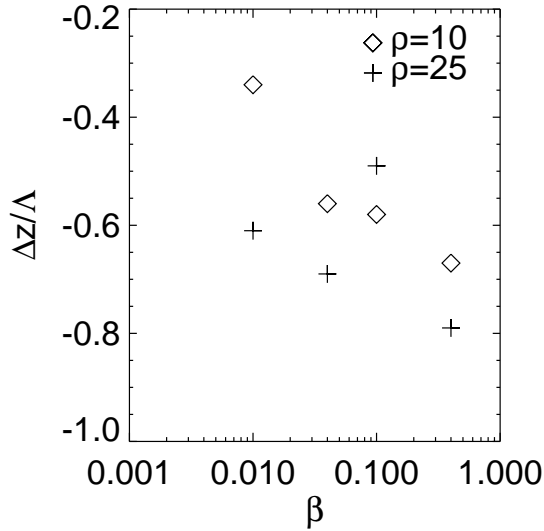}
  \caption{Change in o-point height for model 2. This also includes results from the case where $\beta=0.4$.}
\label{Model2_opoint}
\end{figure}

\begin{figure}[ht]
\centering
\includegraphics[width=8cm]{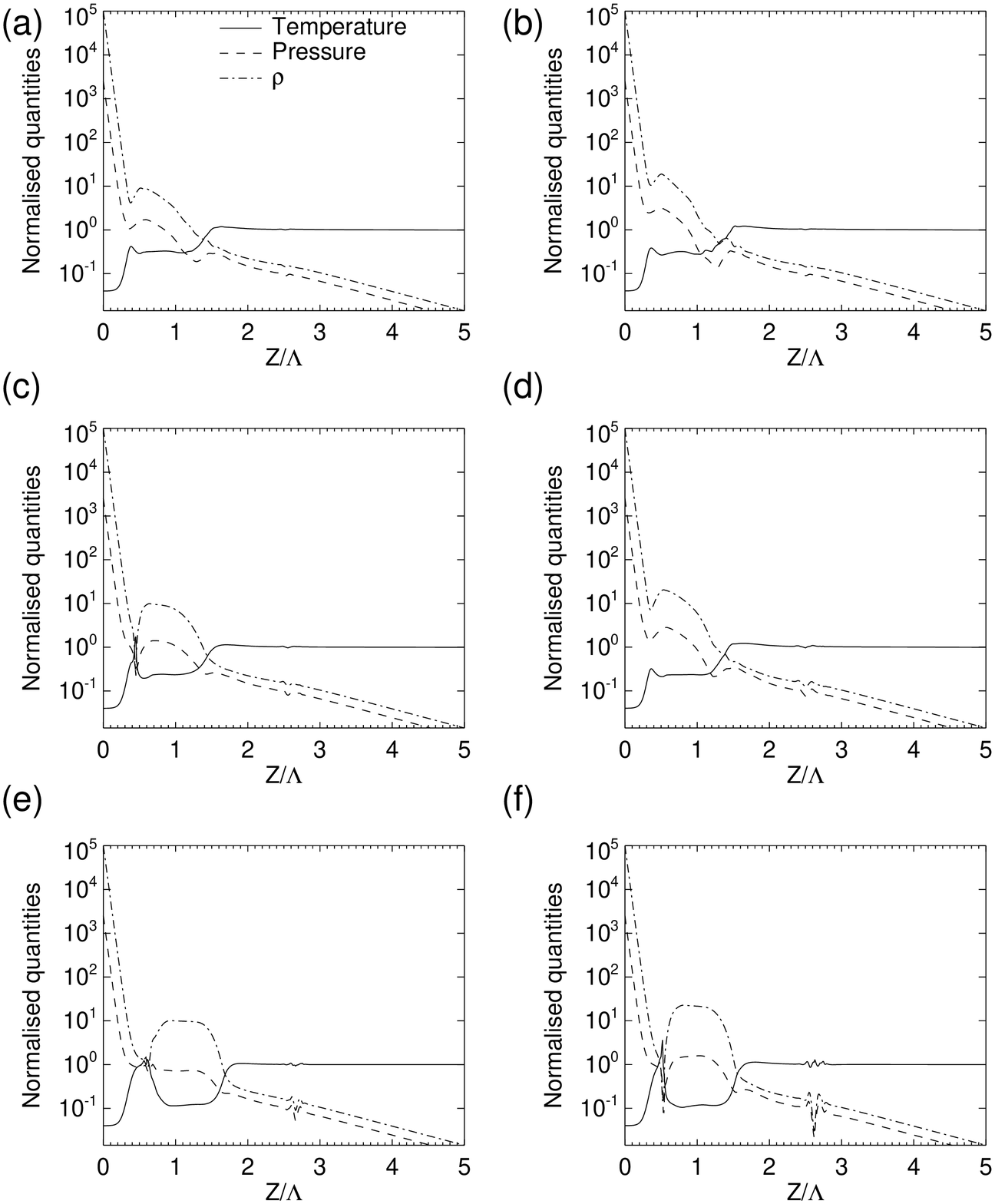}
  \caption{Vertical distribution of the hydrodynamic variables for model 2. The distributions are taken at the horizontal position $x=0\Lambda$.}
\label{Model2_vertical}
\end{figure}

\begin{figure*}[ht]
  \begin{center}
\includegraphics[width=14cm]{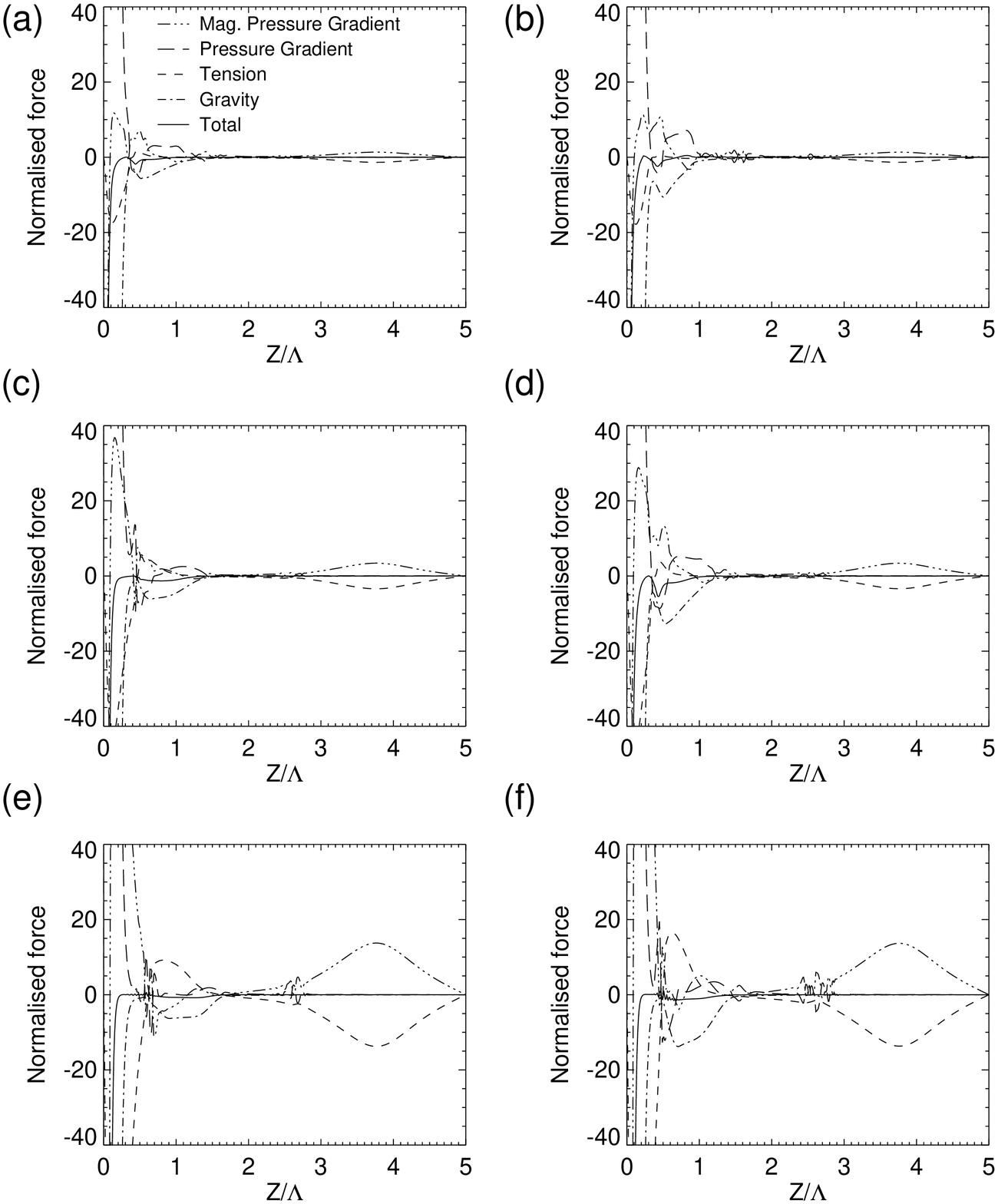}

  \end{center}
  \caption{Force distribution for the six different parameter sets for model 2. The distributions are taken at the horizontal position $x=0\Lambda$.}
\label{Model2_force}
\end{figure*}

\begin{figure}[ht]
  \centering
\includegraphics[width=14cm]{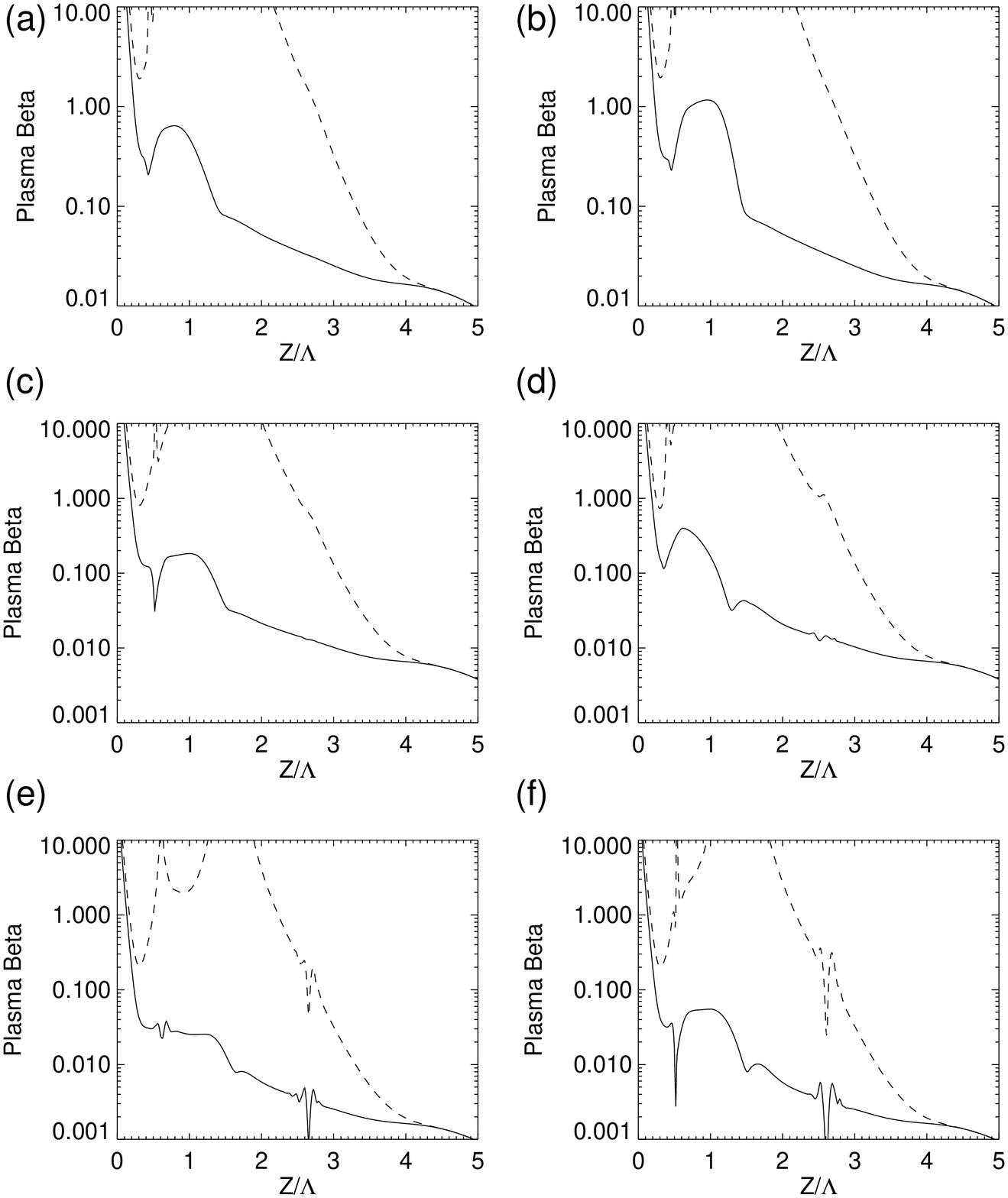}
  \caption{Vertical distribution of plasma $\beta$ for model 2. The solid line shows the plasma $\beta$ for all magnetic field components, and the dashed line shows the plasma $\beta$ for the $B_x$ component, which is related to the amount of tension the magnetic field can exert.}
\label{Model2_beta}
\end{figure}

\begin{figure}
\centering
\includegraphics[width=12cm]{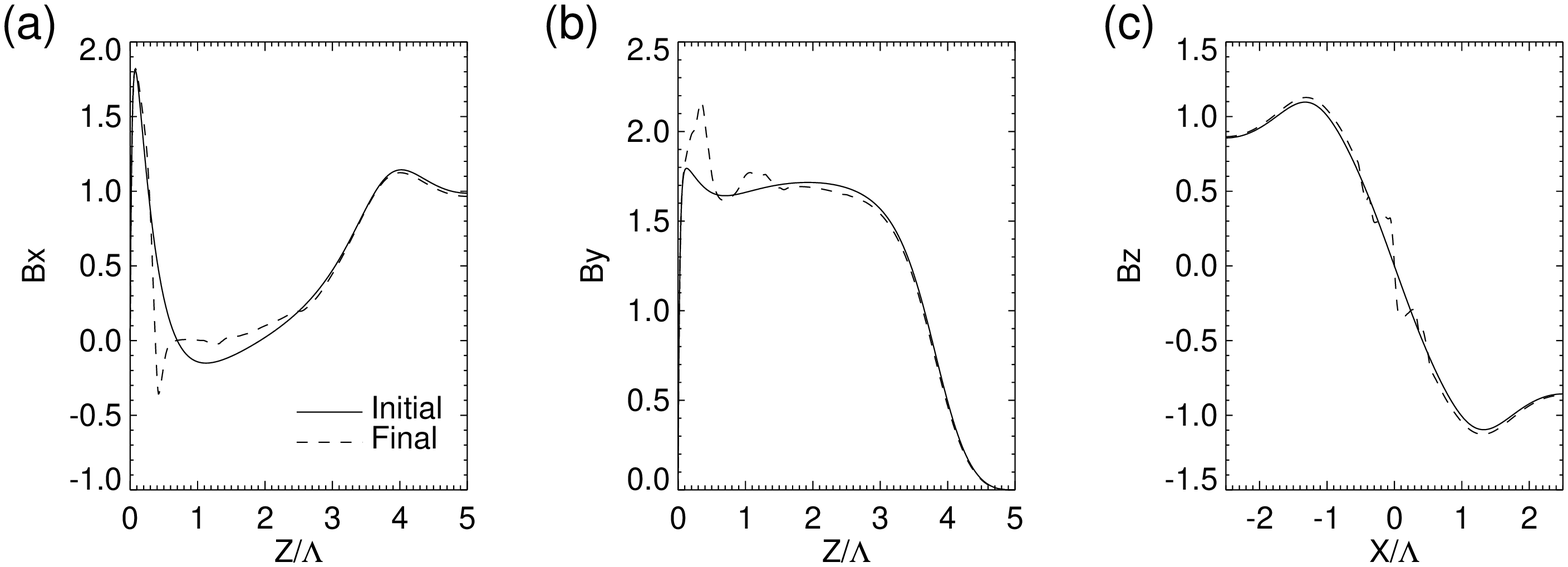}\\
\includegraphics[width=12cm]{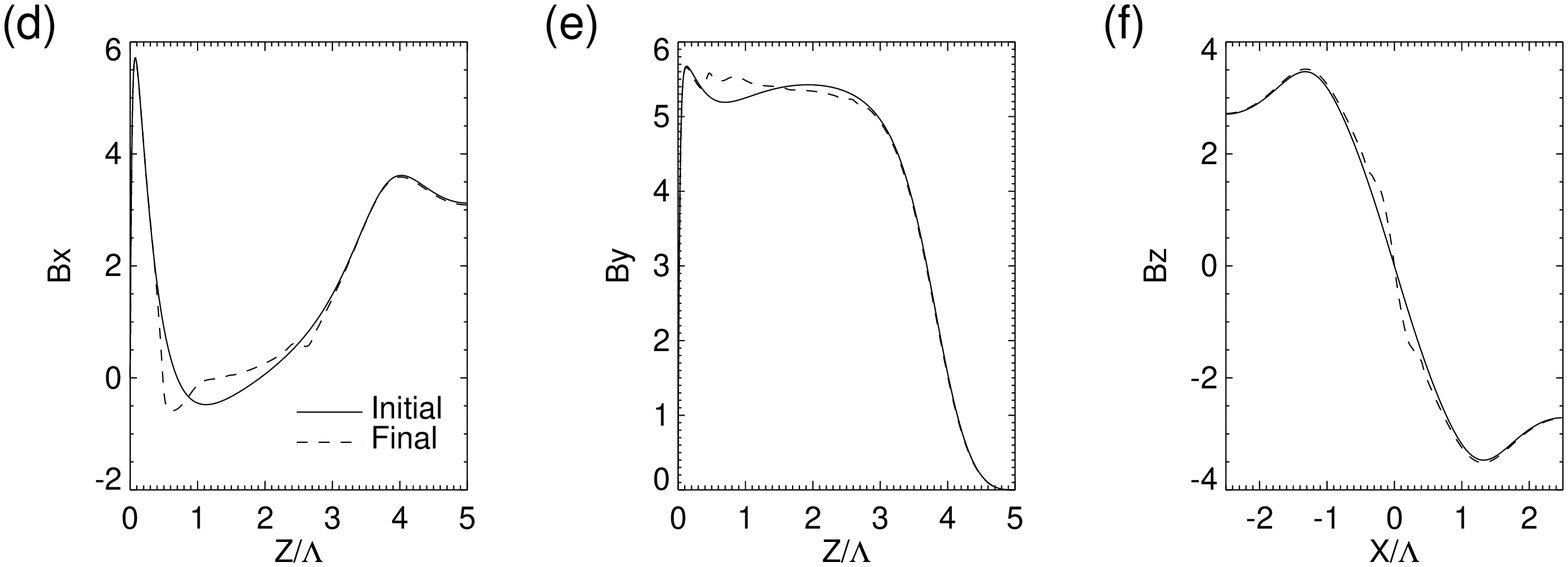}
  \caption{Distribution of the individual components of the magnetic field for model 2 parameter set b (top row) and parameter set f (bottom row). The $B_x$ and $B_y$ distributions are taken at $x=0$ and the $B_z$ distribution is taken at $z=1\Lambda$. }
\label{Model2_mag_dist}
\end{figure}

\begin{figure*}[ht]
  \begin{center}
\includegraphics[width=8cm]{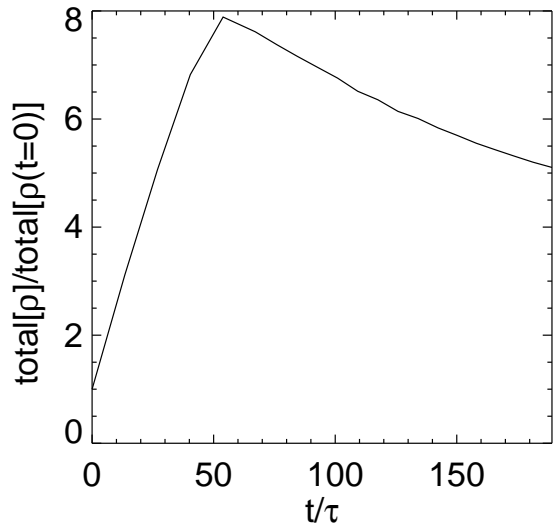}
   
    %%% \FigureFile(width,height){filename}
  \end{center}
  \caption{Temporal evolution of normalised mass in the flux tube for case f of model 2. This figure highlights the drain in mass from the flux tube due to reconnection at the x-point.}
\label{den_loss}
\end{figure*}

Looking at the plasma $\beta$ as shown in Figure \ref{Model1_beta}, only when the plasma $\beta$ of the magnetic field supporting the dense material is less than $0.3$ is it possible to support the prominence material in the $\rho'=25$ case.
If we extrapolate to greater prominence densities, then for support of the material to be possible, it would be necessary that plasma $\beta < 0.1$.
For a prominence with gas pressure of $p=0.3$\,dyn, the required magnetic field strength would be $B>\sqrt{p 8 \pi / 0.1} \sim 8$\,G, which is consistent with the average polar crown prominence magnetic field strength of $\sim 5$\,G \citep{AN2007}.
The relationship between the magnetic field strength and the drop in the height of the flux tube o-point was theoretically analysed by \citet{BLOK2011}, given in Equation (25) of that paper.
However, applying this equation to the results of model 1, it was found that the drop in o-point was grossly overestimated for $\beta=0.4$ and underestimated for $\beta=0.04$.
Therefore, the assumption that gravity can be studied as a small perturbation to the system, as used in the \citet{BLOK2011}, does not apply to the prominences studied in this work.

The mass support we study is 2.5D, therefore the tension term $B_y\partial B_z/\partial y$ cannot be invoked to help support the material.
For model 2, it was not possible to support the dense material due to the large ratio of axial field to horizontal field, with $B_y/B_x \sim 11$.
If variations along the axis of the prominence were allowed, then this axial field could be used for the support. 
Therefore it should be expected that in three dimensions, support of material should be possible for plasma $\beta$ that are closer to unity than the values found in this study.
This would cancel out the factor of four reduction in prominence density for the prominences studied here.
It would be an interesting research topic to extend these simulations to three dimensions. 
It would also be interesting to then extend this work to include Cowling resistivity \citep{BRA1965,COW1957}, as it has been shown to alter the prominence magnetic field over the time scale of the Cowling resistivity \citep{HILL2010}.

In this paper a 2.5D model for a prominence is considered. 
Therefore, the prominence flux rope is not tied to the photosphere at its ends,and will always be unstable to internal kink modes. 
Real world prominences have finite length, and kink instabilities can be suppressed by line-tying effects. 
To gauge the stability of the systems considered here, let us assume that the modelled flux ropes have a length $L_y = 10 \Lambda \approx 550$ Mm. The stability depends on the so-called safety factor, which is defined by $q(A) = B(A)/L_y \oint ds / | \nabla_\perp A |$, where the integral is over a closed contour $A(x,z)=$ constant. 
We find that for model 1 the safety factor increases from about 0.15 at the outer edge of the flux rope to 0.28 on its axis, while for model 2 the safety factor is much larger and nearly constant, $q \approx 2.4$. Kink instability is predicted to occur when $q < 1$, so model 1 is likely to be unstable, but model 2 is stable.

The parameters for model 1 were chosen to make the height of the flux rope axis as large as possible and comparable to the height of observed prominences, $z_{axis} \approx 1.5 \Lambda \approx 83$ Mm, because the axis determines the height of the dips. 
We found that this observational constraint can be satisfied only for relatively high values of the total current $J_y$. 
This leads to highly twisted fields that tend to be kink unstable. We do not believe that real prominences are necessarily so highly twisted. 
However, with the present size of the computational domain and using Equations (4) and (6) we were unable to obtain a weakly twisted flux rope with its axis at sufficiently large height. 
The issue of kink instability is not so important for the present work because we are mainly interested in the magnetic support of the prominence plasma. 
However, in future modelling other expressions for the photospheric flux distribution $A(x,0)$ and/or the axial field $B(A)$ should be explored.

\citet{BLOK2011B} presented the condition for the onset of the continuum convective instability in prominences.
A sufficient condition for the stability of the system is if the Brunt-Vaisala frequency projected onto a flux surface:
\begin{equation}
N^2_{BV,pol}=-\left[\frac{B_{\theta}\cdot \nabla p}{\rho B} \right]\left[\frac{B_{\theta}\cdot \nabla S}{\rho B S} \right]
\end{equation}
is greater than or equal to $0$ ($N^2_{BV,pol} \geq 0$) throughout the plasma.
The models investigated in this paper do not satisfy this condition, with the frequency in the dense prominence material $N^2_{BV,pol}=0$ but $N^2_{BV,pol}<0$ in the regions where the density transitions from the high prominence density to the low coronal density.
As the models studied here relax to a 2.5D MHD equilibria, we know that the systems under study are likely to be stable to $k_y=0$ modes of this instability.
However, it is still possible that the instability could grow for $k_y \neq 0$ modes.

As investigated in Figure 8 of \citet{BLOK2011B}, with larger density or weaker magnetic field, gravitational effects modify the growth rate of the continuum convective instability. 
Therefore, this instability could be occurring in observed prominences resulting in the flows of the prominence material that have been observed \citep[e.g.,][]{KUBO1986, Chae2010}.
It would be very interesting to investigate this instability in terms of the conditions for its onset and the nonlinear evolution of the instability in a prominence geometry as this may provide a different explanation to the other models in the literature so far based on reconnection \citep[e.g.][]{PELO2005, Chae2010, HILL2012b} or condensation formation \citep[e.g.][]{HAE2011, LOW2012a, LOW2012b}.

It must be noted that the free energy of the magnetic fields studied here is rather high, a few times larger than the energy of the potential magnetic field.
It has been shown that coronal magnetic fields with such large free energies would erupt \citep{MOORE2012}.
Here we keep the magnetic field pinned down by using a symmetric boundary at the top of the calculation domain, which allows a stable force-free field to form.
In terms of the study presented here, where the formation of a prominence in magnetic dips is studied, it does not present a large problem, but may be difficult to apply these models to analyse the global stability of the system.
It should be noted that \citet{SVB2012} developed a global NLFFF model of an observed quiescent prominence and found that the model flux rope was stable but close to the limit of stability.
Therefore to appears to be possible to construct stable magnetic equilibria of the type studied in this paper.
Extending the method presented in this paper to study prominences where the magnetic field is based on observed photospheric magnetic field would be an interesting research topic.

\bigskip

The authors thank the staff and students of Kwasan and Hida observatories for their support and comments.
The authors thank the referee for comments and suggestions that helped to improve the presentation of this work.
This work was supported in part by the Grant-in-Aid for the Global COE program ``The Next Generation of Physics, Spun from Universality and Emergence'' from the Ministry of Education, Culture, Sports, Science and Technology (MEXT) of Japan.
This work was supported by the \textit{Hinode}/XRT project under NASA grant NNM07AB07C. \textit{Hinode} is a Japanese mission developed and launched by ISAS/JAXA, with NAOJ as domestic partner and NASA and STFC (UK) as international partners.
It is operated by these agencies in co-operation with ESA and the NSC (Norway).

%Figure 1

%Figure 2

%Figure 3

%Figure 4

%Figure 5

%Figure 6

%FIgure 7

%figure 8

%figure 9

%Figure 10

%Figure 11

%Figure 12

%Figure 13

%Figure 14

%Figure 15

%Figure 16

%Figure 17

%Figure 18

%Figure 19

%Figure 20

%Figure 21

%Figure 22

%Figure 23

%Figure 24

\end{document}